\documentclass[
amsmath,amssymb,twocolumn,aps,
]{revtex4}
\usepackage{graphicx,graphics,color,epsfig}
\usepackage{bm}
\usepackage{dcolumn}
\usepackage{amsmath}
\usepackage{amssymb}
\usepackage{graphicx}
\usepackage{epstopdf}
\usepackage[colorlinks,
anchorcolor=blue,  
linkcolor=blue,       
citecolor=blue]{hyperref}  
\usepackage{natbib}
\usepackage{multirow}  

\newcommand{\be}{\begin{eqnarray}}
\newcommand{\ee}{\end{eqnarray}}

\newcommand{\ba}{\begin{align}}
\newcommand{\ea}{\end{align}}

\makeatletter

\newcommand{\Rmnum}[1]{\expandafter\@slowromancap\romannumeral #1@}
\makeatother

\providecommand{\U}[1]{\protect\rule{.1in}{.1in}}
\begin{document}
\title{Interferometry based on quantum Kibble-Zurek mechanism}
\author{Han-Chuan Kou}
\affiliation{College of Physics, Sichuan University, 610064, Chengdu, People’s Republic of China\\and Key Laboratory of High Energy Density Physics and Technology of Ministry of Education, Sichuan University, 610064,Chengdu, People’s Republic of China}
\author{Peng Li}
\email{lipeng@scu.edu.cn}
\affiliation{College of Physics, Sichuan University, 610064, Chengdu, People’s Republic of China\\and Key Laboratory of High Energy Density Physics and Technology of Ministry of Education, Sichuan University, 610064,Chengdu, People’s Republic of China}

\date{\today}

\begin{abstract}
  We propose an interferometry within the framework of quantum Kibble-Zurek mechanism by exemplifying two prototypical quench protocols, namely the round-trip and quarter-turn ones, on the transverse Ising and quantum $XY$ chains. Each protocol contains two linear ramps that drive the system across quantum critical point twice. The two linear ramps arouse two respective nonadiabatic critical dynamics that are well described by the quantum Kibble-Zurek mechanism. However, in combination, the two critical dynamics can interfere with each other deeply. As an effect of the interference, the dynamical phase is exposed in the final excitation probability, which leads to a quantum coherent many-body oscillation in the density of defects with predictable characteristic period. Thus such an interference is available for direct experimental observations. In the quantum $XY$ model, we show that an interference can also arise from the interplay between two different critical dynamics derived from a critical point and a tricritical point. Furthermore, we demonstrate that the interference influences the dephasing of the excited quasiparticle modes intricately by disclosing a phenomenon of multiple length scales, diagonal and off-diagonal ones, in the defect-defect correlators. It turns out that the dephased result relies on how the diagonal and off-diagonal lengths are modulated by the controllable parameter in a quench protocol.
\end{abstract}

\maketitle


\section{Introduction}
Kibble-Zurek mechanism (KZM) was first proposed in a cosmological setting \cite{Kibble_1976, Kibble_1980}, which reflects the nonadiabatic nature of the critical dynamics in symmetry-breaking phase transitions. Later, its scenario for the creation of topological defects was disclosed to be remarkably adaptable to condensed matter systems being quenched \cite{Zurek_1985, Zurek_1993, Zurek_1996}, which paves the way to experimental tests \cite{Isaac_1991, Mark_1994, Ruutu_1996, Pickett_1996, Monaco_2002, Maniv_2003, Sadler_2006, Weiler_2008, Golubchik_2010, Chiara_2010, Griffin_2012, Chomaz_2015, Yukalov_2015, Zoran_2015}. As the analog in quantum phase transitions, quantum KZM (QKZM) \cite{Damski_2005, Zurek_2005, Polkovnikov_2005, Dziarmaga_2005, Dziarmaga_2010} attracts lots of attention in recent years. Significant progresses in both theory \cite{Levitov_2006, Uwe_2006, Zurek_2007-2, Zurek_2007, Hiroki_2007, Sen_2007, Sen_2008, Sen_2008-2, Sen_2008-3, Polkovnikov_2008, Zurek_2008, Dutta_2009, Zurek_2010, Zurek_2013, Sen_2014, Dutta_2017, Dziarmaga_2019, Zurek_2019, Zurek_2020, Divakaran_2020, Vicari_2020, Kormos_2020, Damski_2020, Dziarmaga_2021} and experiment \cite{Chen_2011, Baumann_2011, Singer_2013, Guo_2014, Braun_2015, Anquez_2016, Carolyn_2016, Guo_2016, Cheng_2016, Zurek_2018, Sachdev_2019, Suzuki_2020, Weinberg_2020, Guo_2020} have been made.

The QKZM is well demonstrated by transverse Ising chain, which can be analyzed thoroughly \cite{Dziarmaga_2005} and emulated by experiment on Rydberg atoms \cite{Sachdev_2019}. Theoretically, its many-body state can be decomposed into different modes (quasiparticles) of two-level quantum subsystems, which facilitates us to calculate quantities efficiently. Studies of quenches have been to a large extent concentrating on the density of defects that is accessible to experiments. The density of defects is counted by summing up the pairs of excited quasiparticles. In a quench process, most modes evolve adiabatically as a superposition of the ground and excited states. While the nearly gapless modes evolve nonadiabatically, which results in eternal defects in the final  many-body state. The nonadiabatic dynamics in the transverse Ising chain is ensured by its energy gap closing near the quantum critical point (QCP) \cite{Lieb_1961, Sachdev_2011}. In practice, given that the system is linearly ramped from its initial ground state across the QCP at a slow but uniform rate controlled by the quench time $\tau_{Q}$, we can parameterize this linear ramp in a smooth form, $\varepsilon(t)=(t-t_{c})/\tau_{Q}$, where $\varepsilon(t)$ is a dimensionless distance from the QCP and $t_c$ marks the time when the QCP is crossed. Generally, QKZM states that the final density of defects scales like $n\varpropto\tau_{Q}^{-d\nu/(1+z\nu)}$, where $d$ is the number of space dimensions, $z$ and $\nu$ are the dynamical and the correlation length exponents respectively \cite{Dziarmaga_2010}. This conclusion means that the quench time sets a fundamental length scale. Specifically, for the quantum Ising chain ($d=1$) in a transverse field, the density of defects is related to the so-called KZ correlation length $\hat{\xi}\propto\sqrt{\tau_{Q}}$ by the relation $n\varpropto \hat{\xi}^{-1}$ since we have $z=1$ and $\nu=1$ \cite{Sen_2009, Pratul_2010, Basu_2012, Basu_2015}.

On the other hand, quantum coherence in many-body systems attracts much attention since it may serve as valuable quantum resources \cite{Alexander_2017, Chitambar_2019}. A recent work shows that the dynamical phase characterized by the length scale, $\propto\sqrt{\tau_{Q}}\ln\tau_{Q}$, can lead to coherent many-body oscillation in the transverse Ising model \cite{Dziarmaga_2022}. But the dynamical phase usually has no effect on the density of defects. In this work, we demonstrate that the density of defects could be influenced by the dynamical phase directly through appropriately designed quench protocols, in which a kind of interference effect is elucidated by several quench protocols applied to the transverse Ising and quantum $XY$ chains. Each protocol consists of two successive linear ramps that conform to the QKZM each \cite{Dziarmaga_2005, Sen_2014, Guo_2020}. We show that an interference occurs as interplay between two successive critical dynamics rendered by the protocol. The interference leads to an exposure of the dynamical phase in the final excitation probability, so that an oscillatory behavior in the density of defects can be observed directly. More intriguingly, we shall further disclose a remarkable phenomenon of multiple length scales in the defect-defect correlator, which can reflect the intricate quantum dephasing of the excited quasiparticle modes in the post-transition state.

The remainder of this paper is organized as follows. Two prototypical quench protocols, the round-trip and quarter-turn ones, are presented in Sec. \ref{round_trip} and \ref{XYchain} respectively. We demonstrate in detail the effect and mechanism of interference by the former and show a flexible way to realize the interference by the latter. In Sec. \ref{transversecorrelator}, we reveal an associate phenomenon of multiple length scales in the defect-defect correlator. At last, a brief summary and discussion is given in Sec. \ref{summary}.


\section{Transverse Ising chain}
\label{round_trip}

In this section, a \emph{round-trip quench protocol} is designed for the transverse Ising chain, which displays the essential elements for realizing the interference effect within the framework of QKZM. We present the details of solution for the quench protocol and elucidate the interference effect through the analysis of the final excitation probability and density of defects. The occurrence of interference is attributed to the mechanism of two successive Landau-Zener transitions. A \emph{reversed round-trip quench protocol} is also discussed at last.

\subsection{The model and round-trip quench protocol}

The transverse field quantum Ising chain reads
\begin{align}\label{H-Ising}
H=-\sum_{j=1}^{N}\left( J\sigma_{j}^{x}\sigma_{j+1}^{x}+g\sigma_{j}^{z}\right),
\end{align}
where $\sigma^{a}_j$ ($a=x,y,z$) are Pauli matrices and the total number of lattice sites $N$ is assumed to be even. The periodic boundary condition, $\sigma^{a}_{N+j}=\sigma^{a}_{j}$, is imposed here. We only consider the ferromagnetic case (i.e. $J>0$) and will henceforth set the reference energy scale as $J=1$. By the Jordan-Wigner transformation,
\begin{align}
  \sigma_{j}^{z}=1-2c_{j}^{+}c_{j},~
  \sigma_{j}^{x}=(c_{j}^{+}+c_{j})\prod_{l<j}(-\sigma_{l}^{z}),\label{Jordan-Winger}
\end{align}
where the spinless fermionic operators satisfy the anticommutation relations, $\{c_{j},c_{l}^{\dagger}\}=\delta_{jl}$ and $\{c_{j}^{\dagger},c_{l}^{\dagger}\}=\{c_{j},c_{l}\}=0$, we can transform the Hamiltonian in Eq. (\ref{H-Ising}) to
\begin{align}
  H=P^{+}H^{+}+P^{-}H^{-},
\end{align}
where $P^{\pm}=\frac{1}{2}\left( 1\pm\prod_{j=1}^{N}\sigma_{j}^{z}\right) $ are projectors on the subspaces with even ($+$) and odd ($-$) parities and the corresponding fermionic Hamiltonians read
\begin{align}
H^{\pm}=\sum_{j=1}^{N}\left(c_{j}c_{j+1}-c_{j}^{\dagger}c_{j+1}+g c_{j}^{\dagger}c_{j}-\frac{g}{2}+\text{H.c.}\right).
\end{align}
In $H^{-}$, the periodic boundary conditions, $c_{N+1}=c_{1}$, and in $H^{+}$, the antiperiodic boundary conditions, $c_{N+1}=-c_{1}$, must be obeyed respectively. It is noteworthy that the ground state exhibits even parity for any nonzero value of $g$ and the parity is a good quantum number.

Because the quench process that we will concentrate on begins in the ground state, we can confine our discussion to $H^{+}$. Adopting the same convention as the one in Ref. \cite{Lieb_1961},
\begin{align}
&c_{j}=\frac{1}{\sqrt{N}}\mathrm{e}^{-i\pi/4}\sum_{q}\mathrm{e}^{iqj}c_{q}, \\
&q=-\pi+\frac{(2j-1)\pi}{N}, ~j\in\{1,\cdots,N\}.
\end{align}
we can rewrite the Hamiltonian as
\begin{align}
H^{+}&=\sum_{q}
\left\{\left(\begin{array}{cc}
c_{q}^{\dagger},c_{-q}
\end{array}\right)
\left(\begin{array}{cc}
\epsilon_{q} &\Delta_{q}\\
\Delta_{q} &-\epsilon_{q}
\end{array}
\right)
\left(\begin{array}{cc}
&c_{q}\\&c_{-q}^{\dagger}
\end{array}\right)+g\right\},
\end{align}
where
\begin{align}
\epsilon_{q}=2(g-\cos q),~\Delta_{q}=2\sin q\label{Bq}.
\end{align}
Next, by the canonical Bogoliubov transformation,
\begin{align}
  c_{q}=u_{q}\eta_{q}-v_{q}\eta_{-q}^{\dagger},
\end{align}
with the coefficients satisfying
\begin{equation}\label{Bogo}
   u_{q}^{2}=\frac{\omega_{q}+\epsilon_{q}}{2\omega_{q}},
   ~v_{q}^{2}=\frac{\omega_{q}-\epsilon_{q}}{2\omega_{q}},
   ~2 u_{q}v_{q} = \frac{\Delta_{q}}{2\omega_{q}},
\end{equation}
we can arrive at the diagonalized form of the Hamiltonian,
\begin{equation}\label{even Hamiltonian}
H^{+}=\sum_{q}\omega_{q}(\eta_{q}^{\dagger}\eta_{q}-\frac{1}{2}),
\end{equation}
where the quasiparticle dispersion reads
\begin{align}
\omega_{q}=\sqrt{\epsilon_{q}^{2}+\Delta_{q}^{2}}.\label{omegaq}
\end{align}

\begin{figure}[t]
\begin{center}
  	\includegraphics[width=3.2in,angle=0]{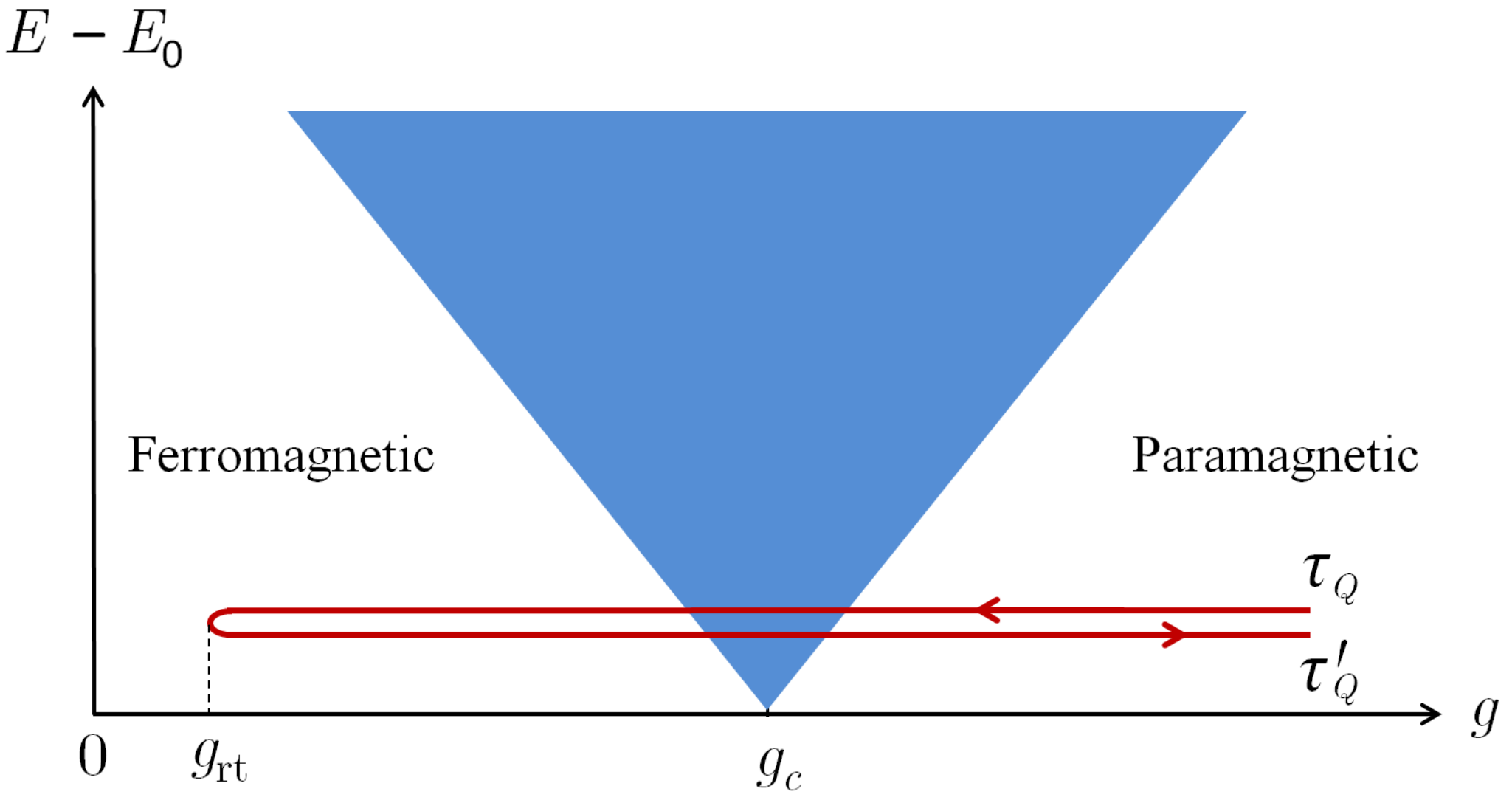}
  \end{center}
  \caption{The round-trip quench protocol consisting of two successive linear ramps. $\tau_Q$ and $\tau'_Q$ are the quench times of the two linear ramps respectively. The shaded area contains all excited energy states. There is an energy gap above the ground state except at the gapless quantum critical point $g_c$.}
  \label{round_trip_quench_scheme}
\end{figure}

In the thermodynamic limit $N\rightarrow\infty$ and at zero temperature, there is a second-order quantum phase transition from a ferromagnetic state ($0<g<1$) with $\mathbb{Z}_{2}$ symmetry breaking to a quantum paramagnetic state ($g>1$) \cite{Sachdev_2011}. The QCP occurs at $g_c = 1$, where the quasiparticle dispersion becomes a linear one, $\omega_q\sim 2|q-q_c|$ with critical quasimomentum $q_c=0$, that is responsible for the dynamical exponent $z=1$. And as depicted in Fig. \ref{round_trip_quench_scheme}, the energy gap between the ground state and the first excited state behaves as $\omega_{0}\varpropto|g-g_c|$, which implies the correlation length exponent $\nu=1$.

Suppose the system is initially prepared in the paramagnetic ground state with all spins polarized up along the transverse field, $|\uparrow,\uparrow,\cdots,\uparrow,\uparrow\rangle$, at a large enough value of $g_i\gg1$. Then a round-trip quench protocol is applied to the system: The system is driven from the paramagnetic to the ferromagnetic regimes and returns to the paramagnetic regime in the end. This protocol contains two linear ramps: one from $g_{i}\equiv g(-\infty)\gg 1$ to $g_{\text{rt}}~(0\leqslant g_{\text{rt}}<1)$ in the first stage and another from  $g_{\text{rt}}$ to $g_{f} \gg 1$ in the second stage. As illustrated in Fig. \ref{round_trip_quench_scheme}, the full procedure of the round-trip quench can be parameterized as \cite{Quan_2010}
\begin{align}
  g\rightarrow g(t) = \left\lbrace
  \begin{array}{cl}
  g_{\text{rt}}-\frac{t}{\tau_{Q}}~~~ &(-\infty < t\leq 0),\\
  g_{\text{rt}}+\frac{t}{\tau'_{Q}}~~~ &(0< t \leq t_{f}),
  \end{array}
  \right.
\end{align}
where $g_{\text{rt}}$ is a turning point, $\tau_{Q}$ and $\tau'_{Q}$ are quench times of the two individual linear ramps respectively, and $t_{f}=(g_{f}-g_{\text{rt}})\tau'_{Q}$ is a reasonably large enough time. In the round-trip quench, the system comes across the QCP twice at $t=-(g_c-g_{\text{rt}})\tau_{Q}$ and $t=(g_c-g_{\text{rt}})\tau'_{Q}$ successively.
Throughout this paper, both $\tau_{Q}$ and $\tau'_{Q}$ are considered to be large enough and $R$ takes moderate values so that QKZM works well in each individual linear quench.

As time evolves, the state of the system gets excited from the instantaneous ground state. We adopt the Heisenberg picture here and assume that the Bogoliubov quasiparticle operators do not change with time, i.e. $i\frac{\mathrm{d}}{\mathrm{d}t}\eta_{q}=0$. The Jordan-Wigner fermions still evolves according to the Heisenberg equation: $i\frac{\mathrm{d}}{\mathrm{d}t}c_{q}=\left[c_{q}, H^{+}\right]$. By a time-dependent Bogoliubov transformation,
\begin{equation}
c_{q}=u_{q}(t)\eta_{q}+v^{*}_{-q}(t)\eta_{-q}^{\dagger},
\end{equation}
we can arrive at the dynamical version of the time-dependent Bogoliubov-de Gennes (TDBdG) equation,
\begin{equation}\label{time-BDG}
i\frac{\mathrm{d}}{\mathrm{d}t}
\left[
\begin{array}{ccc}
u_{q}(t)\\v_{q}(t)
\end{array}
\right] =
\left[
\begin{array}{ccc}
\epsilon_{q}(t)&\Delta_{q}\\\Delta_{q}&-\epsilon_{q}(t)
\end{array}
\right]
\left[
\begin{array}{ccc}
u_{q}(t)\\v_{q}(t)
\end{array}
\right].
\end{equation}
It can be solved exactly by mapping to the Laudau-Zener (LZ) problem. We need to solve this problem for the round-trip quench protocol and calculate the density of defects through the excitation probability in the final state of the system. We shall adopt the long wave approximation appropriately, since only the long wave modes within the small interval, $q\lesssim\frac{1}{\sqrt{\pi\tau_Q}}\ll\frac{\pi}{2}$, can make a contribution and the short wave modes rarely get excited during the quantum phase transition.

At last, in the paramagnetic phase with a large enough value of $g_{f}\gg 1$, the operator of the number of defects $\mathcal{N}_P$ measured by the deviation of spins reduces to the total number of excitations $\mathcal{N}$ approximately \cite{Sen_2014, Guo_2020},
\begin{align}
  \mathcal{N}_P \equiv \frac{1}{2} \sum_{j}(1-\sigma_{j}^{z})\approx\mathcal{N} \equiv \sum_{q}\eta_{q}^{\dagger}\eta_{q}. \label{definition-defect}
\end{align}
One can define the excitation probability as
\begin{equation}
	p_{q}=\langle\eta_{q}^{\dagger}\eta_{q}\rangle=\left|u_{q}(t)v_{q}-v_{q}(t)u_{q}\right|^{2}
,  \label{Exci}
\end{equation}
so that the density of defects is measured by,
\begin{align}
  n = \frac{\langle\mathcal{N}\rangle}{N} = \frac{1}{N}\sum_{q}p_{q},\label{dod}
\end{align}
where $\langle\cdots\rangle$ means the average over the final state of the system. The summation can be replaced by an integral, $\frac{1}{N}\sum_{q}\rightarrow\int_{0}^{\pi}\frac{\mathrm{d} q}{\pi}$, here.

\subsection{The solution for $g_{\text{rt}}=0$}

For simplicity, we dwell on the case of $g_{\text{rt}}=0$ now.

\subsubsection{First stage: \\Linear ramp from $g_{i}\gg 1$ to $g_{\text{rt}}=0$}

In the first stage, we can think the time $t$ varies from $t_{i}=-\infty$ to $0$. The TDBdG equation in Eq. (\ref{time-BDG}) can be transformed into the standard LZ form,
\begin{align}
  &i\frac{\mathrm{d}}{\mathrm{d}\tau}u_{q}(\tau)=-\frac{\tau}{8\tau_{Q}\sin^{2}q}u_{q}(\tau)+\frac{1}{2}v_{q}(\tau),\label{LZ-u}\\ &i\frac{\mathrm{d}}{\mathrm{d}\tau}v_{q}(\tau)=+\frac{\tau}{8\tau_{Q}\sin^{2}q}v_{q}(\tau)+\frac{1}{2}u_{q}(\tau),\label{LZ-v}
\end{align}
where
\begin{equation}
	\tau=4\tau_{Q}\sin q\left(\frac{t}{\tau_{Q}}+\cos q\right) \label{tau}
\end{equation}
is a monotonically increasing parameter. By defining a new time variable $z=\frac{\tau}{2\sqrt{\tau_{Q}}\sin q}e^{i\pi/4}$, we can reduce Eqs. (\ref{LZ-u}) and (\ref{LZ-v}) to the standard Weber equations,
\begin{align}
&i\frac{\mathrm{d}^{2}}{\mathrm{d}z^{2}}u_{q}(z)+(s_{q}-\frac{1}{2}-\frac{z^{2}}{4})u_{q}(z)=0,\\
&i\frac{\mathrm{d}^{2}}{\mathrm{d}z^{2}}v_{q}(z)+(s_{q}+\frac{1}{2}-\frac{z^{2}}{4})v_{q}(z)=0,
\end{align}
where $s_{q}=-i\tau_{Q}\sin^{2}q$. The solutions are expressed in terms of complex parabolic cylinder functions $D_{\nu}(z)$ as \cite{Zener_1932, Frank_2010},
\begin{align}
	v_{q}(z)=&C_{1}D_{-s_{q}-1}(i z)+C_{2}D_{-s_{q}-1}(-i z),\\
	u_{q}(z)=&\frac{e^{i\pi/4}}{\sqrt{\tau_{Q}}\sin q}\left(i\frac{\mathrm{d}}{\mathrm{d}z}+\frac{i z}{2}\right)v_{q}(z)\nonumber\\
	=&\frac{(-1)^{1/4}}{\sqrt{\tau_{Q}}\sin q}\left\{C_{1}D_{-s_{q}}(i z)-C_{2}D_{-s_{q}}(-i z)\right\},
\end{align}
where $C_{1}$ and $C_{2}$ are normalization coefficients. By the initial conditions, $u_{q}(-\infty)=1$ and $v_{q}(-\infty)=0$, we find the two normalization coefficients,
\begin{equation}
	C_{1}=\left.\frac{e^{-i\pi/4}\sqrt{\tau_{Q}}\sin q}{D_{-s_{q}}(i z)}\right|_{t\rightarrow t_{i}=-\infty},~~C_{2}=0. \label{C1C2}
\end{equation}

\subsubsection{Second stage: \\Linear ramp from $g_{\text{rt}}=0$ to $g_{f}\gg1$}

In the second stage, we let the time $t$ vary from $0$ to $\frac{t_{f}}{\tau'_{Q}}\gg1$. The basic TDBdG equation is the same as the one in Eq. (\ref{time-BDG}). However, the equations in LZ form are a little different and read \cite{Zener_1932, Dziarmaga_2005},
\begin{align}
  &i\frac{\mathrm{d}}{\mathrm{d}\tau'}u_{q}(\tau')=+\frac{\tau'}{8\tau'_{Q}\sin^{2}q}u_{q}(\tau')+\frac{1}{2}v_{q}(\tau'),\label{LZ-second_u}\\
	&i\frac{\mathrm{d}}{\mathrm{d}\tau'}v_{q}(\tau')=
          -\frac{\tau'}{8\tau'_{Q}\sin^{2}q}v_{q}(\tau')+\frac{1}{2}u_{q}(\tau'),\label{LZ-second_v}
\end{align}
where $u_{q}(\tau'_{Q})$ and $v_{q}(\tau'_{Q})$ are the Bogoliubov parameters in the second stage and
\begin{equation}
\tau'=4\tau'_{Q}\sin q\left(-\frac{t}{\tau'_{Q}}+\cos q\right)
\end{equation}
is a monotonically decreasing parameter. So Eqs. (\ref{LZ-second_u}) and (\ref{LZ-second_v}) are reduced to
\begin{align}
	&i\frac{\mathrm{d}^{2}}{\mathrm{d}w^{2}}u_{q}(w)+(s'_{q}+\frac{1}{2}-\frac{w^{2}}{4})u_{q}(w)=0,\\
	&i\frac{\mathrm{d}^{2}}{\mathrm{d}w^{2}}v_{q}(w)+(s'_{q}-\frac{1}{2}-\frac{w^{2}}{4})v_{q}(w)=0,
\end{align}
where the time variable becomes $w=\frac{\tau'}{2\sqrt{\tau'_{Q}}\sin q}e^{i\pi/4}$ and $s'_{q}=-i\tau'_{Q}\sin^{2}q$. Likewise, the solution is
\begin{align}
  u_{q}(w)&=C'_{1}D_{-s'_{q}-1}(i w)+C'_{2}D_{-s'_{q}-1}(-i w),\\
  v_{q}(w)&=\frac{\exp^{i\pi/4}}{\sqrt{\tau'_{Q}}\sin q}\left(i\frac{\mathrm{d}}{\mathrm{d}w}+\frac{i w}{2}\right)u'_{q}(w),
\end{align}
where $C'_{1/2}$ are normalization coefficients,
\begin{align}
	C'_{1}=&\frac{D_{-s'_{q}}(-iw_{0})u_{q}^{0}+\frac{v_{q}^{0}\sqrt{\tau'_{Q}}\sin q}{(-1)^{1/4}}D_{-s'_{q}-1}(-iw_{0})}{de(s'_{q}, w_0)},\\
	C'_{2}=&\frac{D_{-s'_{q}}(iw_{0})u_{q}^{0}-\frac{v_{q}^{0}\sqrt{\tau'_{Q}}\sin q}{(-1)^{1/4}}D_{-s'_{q}-1}(iw_{0})}{de(s'_{q}, w_0)},
\end{align}
with $w_{0}=w\left|_{t\rightarrow 0}\right.$ and the denominator of $C'_{1/2}$ reads
\begin{align}
	de(s'_{q}, w_{0})=&\sum_{\alpha=\pm}D_{-s'_{q}-1}(-i \alpha w_{0})D_{-s'_{q}}(i \alpha w_{0}).
\end{align}

\subsubsection{Asymptotic analysis of the solutions}

To reduce the above rigorous solution with elementary functions, we need to apply the asymptotes of $D_{m}(z)$ that are given by \cite{Frank_2010}
\begin{align}
	D_{m}(z)=e^{-z^{2}/4}z^{m},~\forall |\arg(z)|<3\pi/4,
\end{align}
\begin{align}
  D_{m}(z)=&e^{-z^{2}/4}z^{m}-\frac{\sqrt{2\pi}}{\Gamma(-m)}e^{-im\pi}e^{z^{2}/4}z^{-m-1},\nonumber \\
  &~\forall -5\pi/4<\arg(z)<-\pi/4.
\end{align}
We concern the asymptotic solution at two moments: one is $t=0$ (the end of the first stage), another is $t=t_f$ (the end of the second stage).

At the end of the first stage ($t = 0$), the solution valid for $q\lesssim\frac{1}{\sqrt{\pi\tau_Q}}$ is reduced to
\begin{align}
  &u_{q}^{0}\equiv u_{q}(0)=\frac{C_{1}}{\sqrt{\tau_{Q}}\sin q}e^{-3\pi\tau_{Q}\sin^{2}q/4}e^{i\theta_{q}^{u}}, \label{solution-u}\\
  &v_{q}^{0}\equiv v_{q}(0)=\frac{C_{1}\text{sgn}(q)\sqrt{2\pi}}{|\Gamma(1+s_{q})|}
e^{-\pi\tau_{Q}\sin^{2}q/4}e^{-i\theta_{q}^{v}},\label{solution-v}
\end{align}
where coefficient $C_{1}$ fulfills
\begin{equation}
	|C_{1}|^{2}=\tau_{Q}\sin^{2}q~e^{-\pi\tau_{Q}\sin^{2}q/2},
\end{equation}
and the two phase angles are
\begin{align}
  &\theta_{q}^{u}=\tau_{Q}\cos^{2}q+\frac{\tau_{Q}\sin^{2}q}{2}\ln(4\tau_{Q})
  +\arg\left\{\Gamma(1+s_{q}) \right\},\label{theta_u_q}\\
  &\theta_{q}^{v}=\frac{\pi}{4}+\tau_{Q}\cos^{2}q
  +\frac{\tau_{Q}\sin^{2}q}{2}\ln(4\tau_{Q}),\label{theta_v_q}
\end{align}
$\Gamma(x)$ is the gamma function, $\frac{\sqrt{2\pi}}{|\Gamma(1+i x)|}=\sqrt{2\sinh(\pi x)/x}$, and $\arg\{\Gamma(1+i x)\}\approx-\gamma_{E}x$ with Euler constant $\gamma_{E}\approx0.577216$. Please notice that the initial conditions in the second stage are exactly Eqs. (\ref{solution-u}) and (\ref{solution-v}).

At the end of second stage ($t=t_f$), the solution is reduced to
\begin{align}
  u_{q}(t_{f})&=\sqrt{(1-e^{-2\pi \tau'_{Q}\sin^{2}q})(1-e^{-2\pi\tau_{Q}\sin^{2}q})}e^{-i(\theta_{q}^{v}+\phi_{q}^{a})}\nonumber\\
	&+e^{-\pi \tau_{Q} \sin^{2}q}e^{-\pi \tau'_{Q}\sin^{2}q}e^{i(\theta_{q}^{u}-\phi_{q}^{b})},\label{uqtf}\\
  v_{q}(t_{f})&=-e^{-\pi \tau_{Q}\sin^{2}q}\sqrt{1-e^{-2\pi \tau'_{Q}\sin^{2}q}}e^{i(\theta_{q}^{u}+\phi_{q}^{a})}\nonumber\\
	&+e^{-\pi \tau'_{Q} \sin^{2}q}\sqrt{1-e^{-2\pi\tau_{Q}\sin^{2}q}}e^{-i(\theta_{q}^{v}-\phi_{q}^{b})},\label{vqtf}
\end{align}
where $\theta_{q}^{u/v}$ are to be found in Eqs. (\ref{theta_u_q})-(\ref{theta_v_q}) and
\begin{align}
	\phi_{q}^{a}=&\frac{\pi}{4}+\tau'_{Q}\left\{(g_{f}-\cos q)^{2}+\cos^{2}q\right\}+\arg\left\{\Gamma(1+s'_{q}) \right\}\nonumber\\ &+\tau'_{Q}\sin^{2}q\ln\left\{4\tau'_{Q}(g_{f}-\cos q)\cos q \right\}, \label{phi_a}\\
	\phi_{q}^{b}=&\tau'_{Q}\left\{(g_{f}-\cos q)^{2}-\cos^{2}q\right\}+\tau'_{Q}\sin^{2}q\ln\frac{g_{f}-\cos q}{\cos q},  \label{phi_b}
\end{align}
are another two phase angles produced by the second stage of the quench process.

\subsection{Interference effect for $g_{\text{rt}}=0$}

\subsubsection{Final excitation probability}

Following the classical works \cite{Zurek_2005,Dziarmaga_2005}, we obtain the excitation probability at $t=0$,
\begin{align}
	p_{q}^{0}\equiv p_{q}(0)\approx e^{-2\pi\tau_Q q^2}. \label{excitation_0}
\end{align}
It reaches its peak at $q^{*}=\pi/N\sim0$. In the calculation, we have made the substitutions, $\sin^{2}q\approx q^{2}$ and $\cos^{2}q\approx 1-q^{2}$, according to the long wave approximation \cite{Sen_2014}. It is clear that the phase angles in Eqs. (\ref{theta_u_q}) and (\ref{theta_v_q}) do not matter at present.

\begin{figure}[t]
\begin{center}
  	\includegraphics[width=3.4in,angle=0]{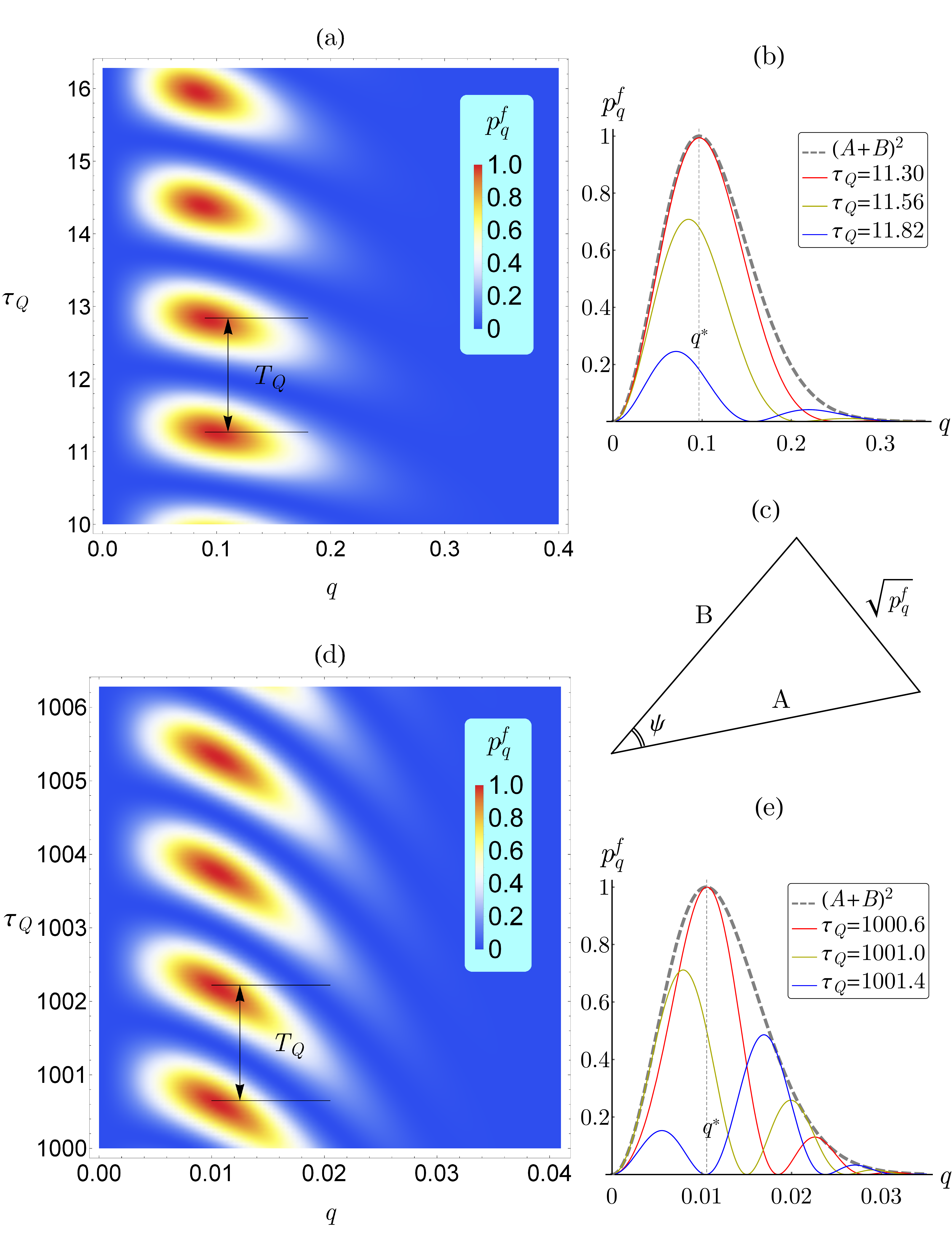}
  \end{center}
  \caption{Analysis of the final excitation probability $p_q^f$ at $R=1$. (a) and (d): Density plot of $p_q^f$. (b) and (e): Peak structure of $p_q^f$ at several selected values of $\tau_Q$. (c): Graph representation of Eq. (\ref{excitation_f}). The density plots illustrate the oscillatory behavior of $p_q^f$ along the quench time with a period $T_{Q}=\pi/2$ according to Eq. (\ref{T_Q}). The gray dashed lines in (b) and (e) denote the upper bound of the peaks, $(A+B)^2$, whose position is located at $q^{*}=\sqrt{\ln 2/(2\pi\tau_Q)}$ for $R=1$.}
  \label{pq}
\end{figure}

Next, by Eq. (\ref{Exci}), we work out the final excitation probability after the full round-trip quench as
\begin{align}
	p_{q}^{f}\equiv p_{q}(t_f)=A^{2}+B^{2}-2AB\cos\psi, \label{excitation_f}
\end{align}
where
\begin{align}
  &A=e^{-\pi q^{2}\tau_{Q}}\sqrt{1-e^{-2\pi q^{2} \tau_{Q} R}}, \label{equation_A}\\
  &B=e^{-\pi q^{2} \tau_{Q} R}\sqrt{1-e^{-2\pi q^{2}\tau_{Q}}},\label{equation_B}\\
  &\psi=\theta_{q}^{u}+\theta_{q}^{v}+\phi_{q}^{a}-\phi_{q}^{b}. \label{equation_psi}
\end{align}
For convenience, we have introduced the ratio,
\begin{align}
  R=\frac{\tau'_Q}{\tau_Q}. \label{R}
\end{align}
The presence of the total dynamical phase $\psi$ in $p_q^f$ undoubtedly manifests an interference effect. To see it through, we display the density plots of $p_{q}^{f}$ in Fig. \ref{pq}. In it, we observe that an array of peaks with characteristic quasimomentum $q^{*}\sim\tau_Q^{-1/2}$ periodically lines up along $\tau_Q$ direction. Two features can be inferred. First, $p_{q}^f$ takes values between lower and upper bounds,
\begin{align}
  (A-B)^2 \leq p_{q}^{f} \leq (A+B)^2. \label{bounds}
\end{align}
The characteristic quasimomentum $q^{*}$ is defined as the peak's position of the upper bound, whose value can be numerically solved from the equation,
\begin{eqnarray}
  \frac{1+\sqrt{2}e^{\pi(R-1)\tau_Q q^2}}{R\sqrt{R}+e^{\pi(R-1)\tau_Q q^2}}=\frac{e^{\pi(R+1)\tau_Q q^2}}{1+R}.
\end{eqnarray}
While for $R=1$, we can solve it analytically and get $q^{*}=\sqrt{\frac{\ln2}{2\pi\tau_Q}}$.
\begin{figure*}[t]
  \centering
  \includegraphics[width=6.8in,angle=0]{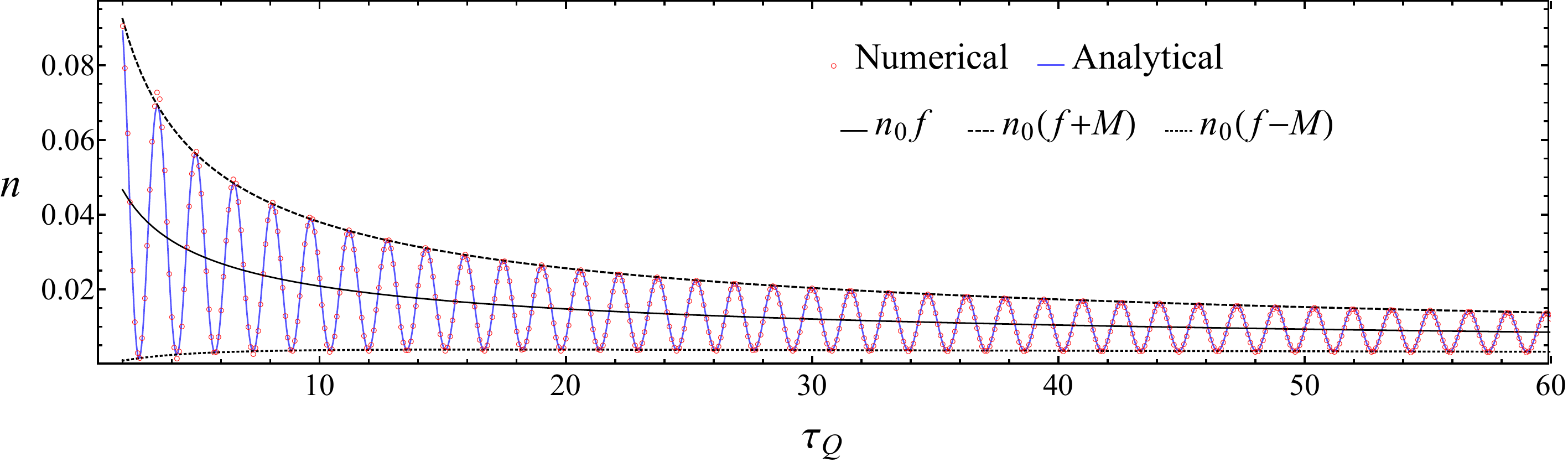}
  \caption{Coherent oscillation of the density of defects. The blue oscillatory curve is the analytical solution in Eq. (\ref{simp-dd}), which is in very agreement with the numerical one (red circles) obtained by TDBdG equation in Eq. (\ref{time-BDG}). The black solid curve denotes the non-oscillatory part, $n_0 f$. The dashed and dotted curves denote the upper and lower bounds, $n_0(f+M)$ and $n_0(f-M)$, of the oscillation. Please notice we choose the parameter $R=1$ here, thus the period of oscillation is $T_Q = \pi/2$.}
  \label{defect-density}
\end{figure*}
Second, the period of oscillation along $\tau_Q$ direction can be worked out. For large enough $\tau_Q$, we can reduce the total dynamical phase to
\begin{align}
  \psi&=\frac{\pi}{2}+2(1+R)\tau_{Q}+q^{2}\tau_{Q}\left\{(1+R)\ln\tau_{Q}\right. \nonumber\\
  &~~~~~+\left.(1+R)(\ln4+\gamma_{E}-2) + R\ln R\right\}. \label{app-psi}
\end{align}
It contains two important lengths, $\tau_Q$ and $\tau_Q\ln\tau_Q$. The former determines the period,
\begin{equation}
  T_{Q}=\lim_{\tau_Q\rightarrow\infty}\frac{2\pi}{(1+R)\{2+(q^{*})^{2}\ln\tau_{Q}\}}
       =\frac{\pi}{1+R}, \label{T_Q}
\end{equation}
because only the modes around the peaks at $q^{*}$ makes contributions and the value of $q^{*}$ is as small as $\sim\tau_Q^{-1/2}$. The same oscillatory behavior can also be measured by the variable $\tau'_Q$ and the corresponding period becomes
\begin{equation}
  T'_{Q}=\frac{T_Q}{R}=\frac{\pi}{R(1+R)}. \label{Tp_Q}
\end{equation}

In $\psi$, the term with length $\tau_{Q}\ln\tau_{Q}$ implies dephasing of the excitation modes when $\tau_Q$ is large enough. And due to the interference, the defect-defect correlator would be affected by this length in a more profound way. We will look into this issue in Sec. \ref{transversecorrelator}.

\subsubsection{Oscillatory density of defects}

As a more easy way to observe the effect of interference, we work out the final density of defects as
\begin{align}
  n= n_0 \left\{f + \sum_{i=1}^{3} M_{i}\cos\left(2\pi\frac{\tau_Q}{T_Q}+\delta_{i}\right)\right\}, \label{n_text}
\end{align}
in which
\begin{align}
  n_{0}=\frac{1}{2\pi\sqrt{2\tau_{Q}}} \label{n0}
\end{align}
is a QKZM factor matching the result of usual one-way quench \cite{Dziarmaga_2005},
\begin{align}
  f=1+\frac{1}{\sqrt{R}}-\frac{2}{\sqrt{1+R}} \label{f}
\end{align}
is a factor without oscillation, $M_{i}$'s and $\delta_{i}$'s are oscillation amplitudes and phases whose expressions are to be found in Appendix \ref{Appendix_A}. This result can be abbreviated to
\begin{eqnarray}
  n = n_0 \left\{f + M \cos\left(2\pi\frac{\tau_Q}{T_Q}+\delta\right)\right\}. \label{simp-dd}
\end{eqnarray}
To distinguish the contribution of each term, we write down
\begin{eqnarray}
  n_{0}f=\frac{1}{2\pi\sqrt{2\tau_{Q}}} + \frac{1}{2\pi\sqrt{2\tau'_{Q}}} -\frac{1}{\pi\sqrt{2(\tau_{Q}+\tau'_{Q})}}. \label{n0f}
\end{eqnarray}
So it is clear to see that the former two terms in Eq. (\ref{n0f}) are individual contributions from the two critical dynamics of linear quenches and the last term is a nonoscillatory part of interplay between the two  critical dynamics. While the oscillatory part of interplay between the two  critical dynamics is embedded in the cosinusoidal terms in Eq. (\ref{n_text}) or the one in Eq. (\ref{simp-dd}). In the large $\tau_Q$ limit, the amplitude $M$ becomes
\begin{equation}
  \lim\limits_{\tau_{Q},\tau'_{Q}\rightarrow\infty}M=\frac{2\pi\tau_{Q}\sqrt{2\pi\tau'_{Q}}}{(\tau_{Q}\ln\tau_{Q}+\tau'_{Q}\ln\tau'_{Q})^{3/2}} \label{limM}
\end{equation}
asymptotically. When $R=1$, we have $\lim_{\tau_{Q}\rightarrow\infty}M\sim(\ln\tau_Q)^{-3/2}$. This result means that the amplitude $M$ (or $M_i$'s in Eq. (\ref{n_text})) plays the role of dephasing factor actually \cite{Dziarmaga_2022}.

The oscillatory density of defects for $R=1$ and $2<\tau_Q<60$ is exemplified in Fig. \ref{defect-density}. We see the formula in Eq. (\ref{simp-dd}) obtained by long wave approximation is in good agreement with the numerical solution of the TDBdG equation in Eq. (\ref{time-BDG}). This intriguing result is in contrast to the traditional case of one-way quench \cite{Dziarmaga_2005}.

\subsubsection{Mechanism of interference: \\two successive Landau-Zener transitions}

\begin{figure}[t]
\begin{center}
  	\includegraphics[width=3.3in,angle=0]{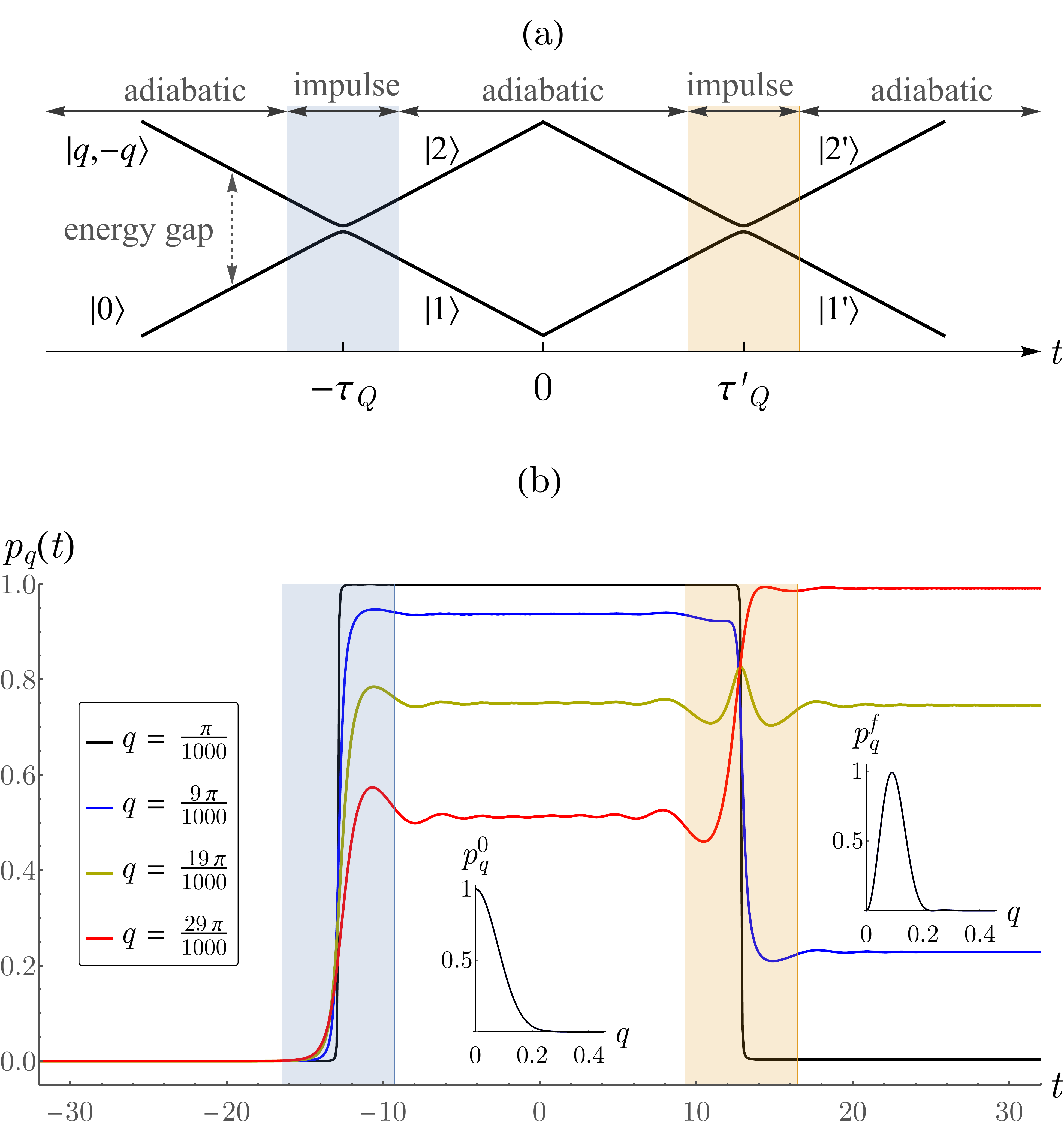}
  \end{center}
  \caption{(a) Two successive Landau-Zener transitions for $(q,-q)$ modes in the round-trip quench process. The full process can be divided into three adiabatic and two impulse stages approximately. (b) Evolution of the excitation probability. The final excitation probability is output for large enough time $t$. The two insets show a shift of peak of the excitation probability $p_q(t)$ evolving from $p_q^0$ to $p_q^f$. We have selected the parameters $R=1$ and $\tau_Q=12.87$ in this demonstration.}
  \label{interference_excitation}
\end{figure}

Now we explain why the quantum dynamical phase can result in an interference in the round-trip quench protocol. In fact, the interference can be attributed to a mechanism basing on a theory of two successive Landau-Zener transitions. As illustrated in Fig. \ref{interference_excitation} (a), the full round-trip quench process can be divided into three adiabatic and two impulse stages approximately \cite{Dziarmaga_2010}. The system is driven across QCP twice, i.e. the system undergoes nonadiabatic transitions in the two impulse regimes as represented by the shaded areas. To get a rough but clear picture, we discuss the theory in an intuitive way below.

First of all, different pairs of quasiparticles get excited independently, thus we can focus on the single mode problem. In the initial state, the $q$ mode is empty and we label the state by $|0\rangle$. With time increasing, the $q$ mode state that is labelled by $|q,-q\rangle=c_q^{\dagger}c_{-q}^{\dagger}|0\rangle$ is involved. Then we can follow the evolution of the two states only concerning the $q$ mode. According to the standard Landau-Zener transition theory, the two states after the first nonadiabatic transition can be written down as \cite{Li_2021}
\begin{align}
  |1\rangle \sim &\sqrt{1- p_{q}^{0}}~e^{i\theta_{1}}|0\rangle+ \sqrt{p_{q}^{0}}~e^{i\theta_{2}}|q,-q\rangle, \\
  |2\rangle \sim &\sqrt{p_{q}^{0}}~e^{-i\theta_{2}}|0\rangle-\sqrt{1- p_{q}^{0}}~~e^{-i\theta_{1}}|q,-q\rangle,
\end{align}
where $p_{q}^{0}=e^{-2\pi\tau_{Q}q^2}$, $\theta_{1/2}$ are nonzero phases determined by the quench time $\tau_Q$. Next, through the second nonadiabatic transition, the two sates turn into
\begin{align}
  |1'\rangle \sim &\sqrt{1- p'_{q}}~e^{i\phi_{1}}|1\rangle+ \sqrt{p'_{q}}~e^{i\phi_{2}}|2\rangle, \\
  |2'\rangle \sim &\sqrt{p'_{q}}~e^{-i\phi_{2}}|1\rangle-\sqrt{1- p'_{q}}~e^{-i\phi_{1}}|2\rangle,
\end{align}
where $p'_{q}=e^{-2\pi\tau'_{Q}q^{2}}$, $\phi_{1/2}$ are another nonzero phases determined by the quench time $\tau'_Q$.

No interference occurs after the first nonadiabatic transition, since the excitation probability at this moment reads
\begin{align}
  p_{q}^{0}=|\langle q,-q|1\rangle|^{2}=e^{-2\pi\tau_{Q}q^2},
\end{align}
which exhibits a Gaussian peak centered at $q^{*}=0$ and no any information of phase remains. However, an interference is inevitable after the second nonadiabatic transition, because the final excitation probability contains a nonzero phase and reads
\begin{align}
  p_{q}^{f}&=|\langle q,-q|1'\rangle|^{2}\nonumber\\
           &=A^{2}+B^{2}-2AB\cos(\theta_{1}+\theta_{2}+\phi_{1}-\phi_{2}).
\end{align}
This concise result is actually the same as that in Eq. (\ref{excitation_f}). It is easy to see that the phases, $\theta_{1/2}$ and $\phi_{1/2}$, mimic the former ones, $\theta_{q}^{u/v}$ and $\phi_{q}^{a/b}$, in Eq. (\ref{equation_psi}) faithfully.

We have also calculated numerically the evolution of the excitation probability $p_q(t)$ to observe in detail how the two successive Landau-Zener transitions influence the final output. The numerical results for a system with size $N=1000$ are illustrated in Fig. \ref{interference_excitation} (b). We can observe that the mode $q=\pi/N$ approaches the saturate value $1$ after the first linear ramp and drops precipitously down to $0$ after the second linear ramp. And the modes $q\sim\tau_Q^{-1/2}$ indeed have the chance to be excited with a higher probability.

In short, the occurrence of interference can be attributed to the fact that the system gets excited twice in the whole quench process. But one may doubt why there is no such interference effect in the quench process with the transverse field being ramped from $g=\infty$ to $g=-\infty$ in the transverse Ising model \cite{Sen_2007, Zurek_2016, Guo_2020}, since the systems also get excited twice. The answer lies in that both the first and secondary excitations must involve the same modes, e.g. the ones fulfilling the condition, $q\gtrsim q_{c}=0$, in our case. While in previous studies, the first and secondary excitations agitate the modes near $q_{c}= 0$ and $q_{c}= \pi$ independently, thus there is no chance for such an interference to occur.

\subsection{Interference effect for $0<g_{\text{rt}}\leq 1$}

Now we free the turning point from $g_{\text{rt}}=0$ to $0<g_{\text{rt}}\leq1$. We skip the details of deduction since it is not much different from the previous one.

\begin{figure}[t]
  \begin{center}
	\includegraphics[width=3.38in,angle=0]{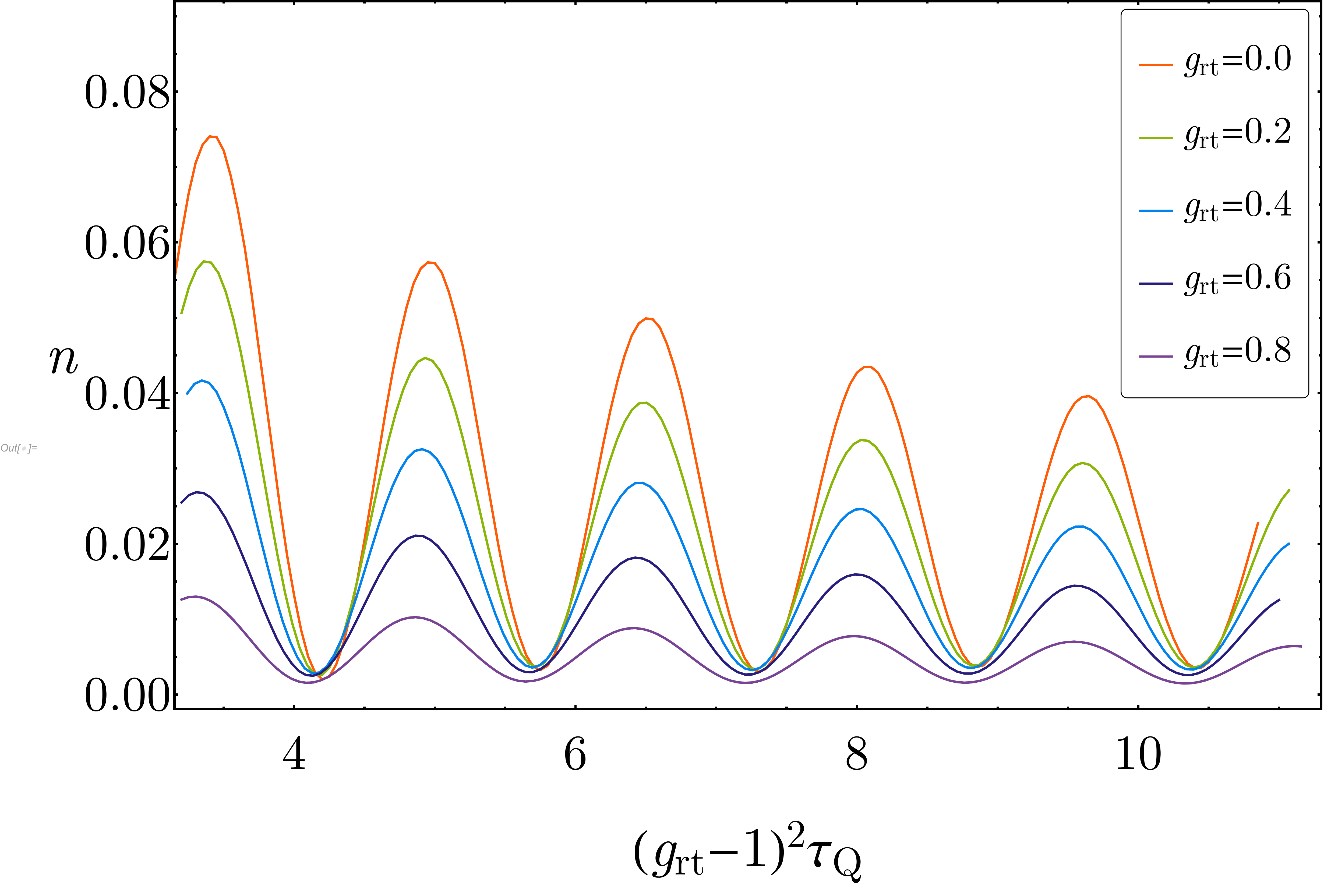}
  \end{center}
  \caption{Density of defects in scaled quench time $(g_{\text{rt}}-1)^{2}\tau_{Q}$. The curves are numerical solutions of the TDBdG equation in Eq. (\ref{time-BDG}) with parameters $R=1$ and $g_{\text{rt}}=0.0,~0.2,~0.4,~0.6,~0.8$. The repeated peaks and troughs justify the formula of period in Eq. (\ref{period-g}).}
    \label{density-scale-g}
\end{figure}

The final excitation probability and density of defects can still be expressed by Eq. (\ref{excitation_f}) and Eq. (\ref{simp-dd}). But the total dynamical phase changes to
\begin{align}
	\psi&=\frac{\pi}{2}+2(g_{\text{rt}}-1)^{2}(1+R)\tau_{Q}+q^{2}\tau_{Q}\left(  R\ln R +\right. \nonumber\\
	&\left.  (1+R)\left[2(g_{\text{rt}}-1)+\ln\{4\tau_{Q}(g_{\text{rt}}-1)^{2}\}+\gamma_{E}\right] \right). \label{psi_gt}
\end{align}
As a consequence, the period of oscillation changes to
\begin{align}
  T_{Q}=\frac{\pi}{(g_{\text{rt}}-1)^{2}(1+R)}. \label{period-g}
\end{align}
The oscillation amplitude of the density of defects also behaves as $\lim_{\tau_{Q}\rightarrow\infty}M\sim(\ln\tau_Q)^{-3/2}$ asymptotically. The oscillatory density of defects in a scaled quench time, $(g_{\text{rt}}-1)^{2}\tau_{Q}$, is numerically solved for several selected values of $g_{\text{rt}}$ and illustrated Fig. \ref{density-scale-g}. The repeated peaks and troughs are approximately located at the same positions, which justify the formula of period in Eq. (\ref{period-g}).

To observe the oscillation more obviously, one should avoid putting the turning point too close to the critical point ($g_{\text{rt}}\rightarrow g_c=1$), because the period goes to infinity and the amplitude disappears in this limit according to Eqs. (\ref{period-g}) and (\ref{limM}) respectively. Nevertheless, there should still be an interference in this case according to the previous discussion basing on the two-step Landau-Zener transitions. But the analytics is a little different. We need to make use of the asymptote of $D_m(z)$,
\begin{equation}
	\lim\limits_{z\rightarrow 0}D_{m}(z)=2^{m/2}\frac{\sqrt{\pi}}{\Gamma(\frac{1}{2}-\frac{m}{2})},
\end{equation}
to get the final excitation probability containing an interference term, $p_{q}^{f} \sim \cos\psi$, where the total dynamical phase turns out to be (Please see details in Appendix \ref{turningpoint})
\begin{align}
  \psi=&\frac{\pi}{2}+\sum_{a}^{\frac{1}{2},1}~\sum_{T}^{\tau_{Q},\tau'_{Q}}(-1)^{2a}\arg\left\lbrace\Gamma(a-\frac{i}{2} T q^{2}) \right\rbrace\nonumber\\
  \approx & \frac{\pi}{2}+\frac{1}{2}\left\lbrace\gamma_{E}+\gamma_{d}\left(\frac{1}{2}\right)\right\rbrace(1+R)\tau_{Q}q^{2}, \label{exci-critical}
\end{align}
where $\gamma_{d}(\frac{1}{2})\approx -1.96351$ and $\gamma_{d}(x)$ is the digamma function. Comparing the total dynamical phase in Eq. (\ref{exci-critical}) with that in Eq. (\ref{app-psi}), we find the term like $2(1+R)\tau_Q$ does not appear in Eq. (\ref{exci-critical}). Thus the final density of defects will behave like, $n\sim\tau_{Q}^{-1/2}$, and there is no oscillation any longer. We have also numerically investigated the case with $g_{\text{rt}}=g_c=1$ on a lattice as large as $N=10000$ and confirmed a QKZM factor without oscillation, $n=0.053\tau_{Q}^{-0.495}\sim\tau_{Q}^{-1/2}$, in the final density of defects. Our observation is consistent with a previous study of a similar system by setting the turning point at its critical point, where an interference term also emerges but no oscillation in the density of defects was found \cite{Dutta_2009}.

\subsection{Reversed round-trip quench protocol}

\begin{figure}[t]
  \begin{center}
    \includegraphics[width=3.2in,angle=0]{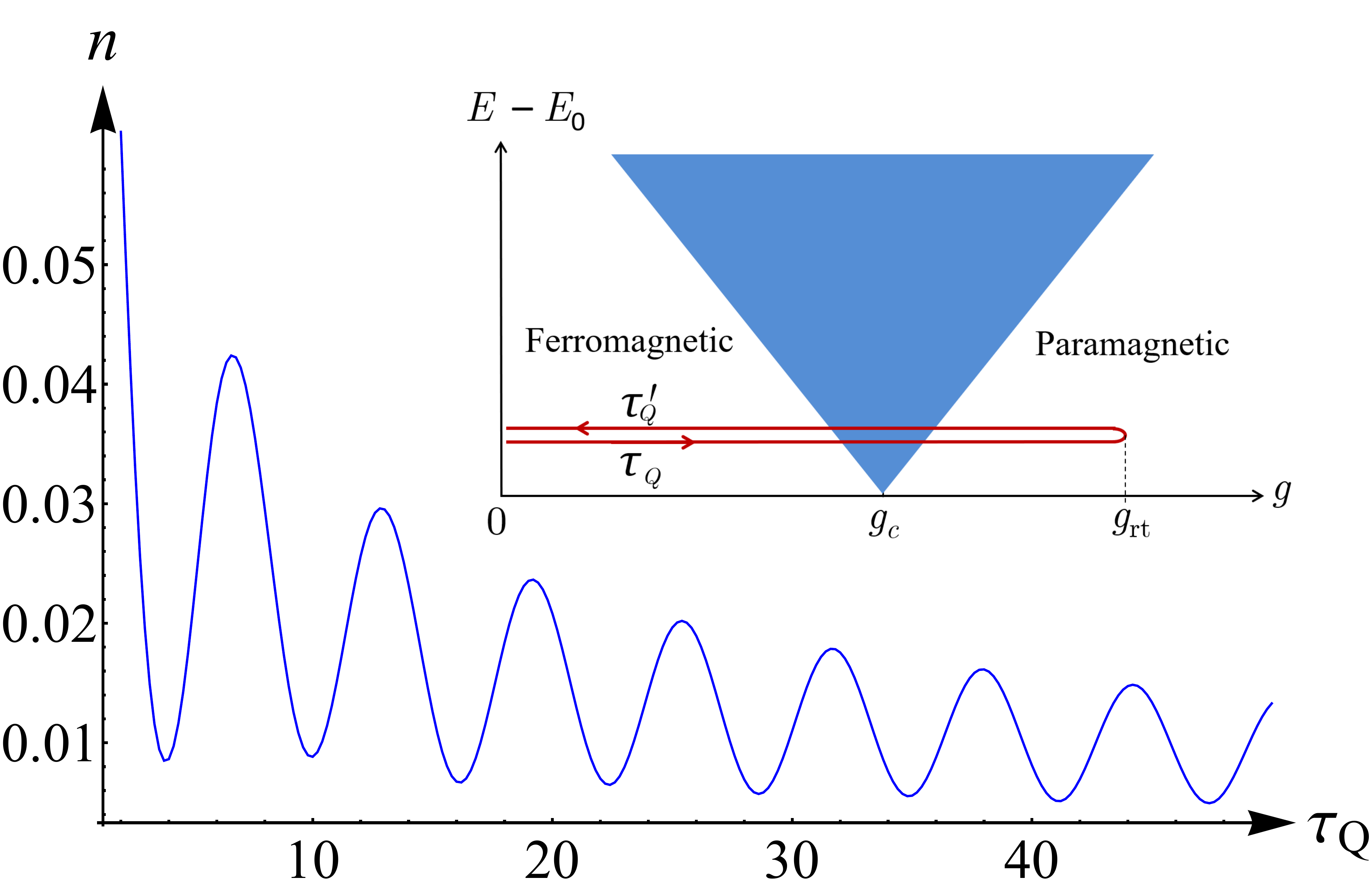}
  \end{center}
  \caption{Density of defects after the reversed round-trip quench protocol is applied to the transverse Ising chain with parameters $R=1$ and $g_{\text{rt}}=1.5$. The period of oscillation is $T_{Q}=2\pi$. The inset shows the reversed round-trip quench protocol.}
  \label{density-reverse}
\end{figure}

In the round-trip quench protocol above, both the starting point and ending point are in the paramagnetic phase. We can also consider a reversed round-trip quench protocol (the inset in Fig. \ref{density-reverse}) for the same Hamiltonian in Eq. (\ref{H-Ising}). The reversed protocol is parameterized as
\begin{equation}
  g\rightarrow \bar{g}(t)=\left\lbrace
  \begin{array}{ccc}
    g_{\text{rt}}+\frac{t}{\tau_{Q}}~ &(-g_{\text{rt}}\tau_{Q} < t\leq 0),\\
    g_{\text{rt}}-\frac{t}{\tau'_{Q}}~ &(0< t \leq g_{\text{rt}}\tau'_{Q}),\\
  \end{array}
  \right.
\end{equation}
where $g_{\text{rt}}$ is the turning point. The initial time is set at $t=-g_{\text{rt}}\tau_{Q}$, where the system is a classical Ising model with zero transverse field. In the first stage, the dynamics is ramped up from the ferromagnetic phase to the paramagnetic one.
At $t=0$, the transverse field reaches the turning point, $g(0)=g_{\text{rt}}>1$, where the first nonadiabatic process has been sufficiently accomplished, i.e. we would get a usual result falling in the QKZM. Then, in the second stage, the transverse field is linearly ramped down and the system finally returns back to the classical Ising model, i.e. the limit of the ferromagnetic phase. In this limit, the kinks along the $x$ axis play the role of defects and the number of defects $\mathcal{N}_F$ matches the number of excitations $\mathcal{N}$ exactly \cite{Dziarmaga_2005},
\begin{align}
  \mathcal{N}_F \equiv \frac{1}{2} \sum_{j}(1-\sigma_{j}^{x}\sigma_{j+1}^{x}) =\mathcal{N} \equiv \sum_{q}\eta_{q}^{\dagger}\eta_{q}. \label{IsingKink-x}
\end{align}

In fact, the two definitions of the number of defects in the limit of paramagnetic phase $\mathcal{N}_P$ and the limit of ferromagnetic phase $\mathcal{N}_F$ are dual to each other. One can see this clearly by introducing the dual transformation \cite{Kogut_1979}, $\mu_{j}^{z}=\sigma_{j}^{x}\sigma_{j+1}^{x}$ and $\mu_{j}^{x}=\prod_{k<j}\sigma_{k}^{z}$, which leads to the mapping of defects,
\begin{align}
  \frac{1}{2} \sum_{j}(1-\sigma_{j}^{z})~\leftrightarrow~\frac{1}{2}\sum_{j}(1-\mu_{j}^{x}\mu_{j+1}^{x}),\\
  \frac{1}{2}\sum_{j}(1-\sigma_{j}^{x}\sigma_{j+1}^{x})~\leftrightarrow~\frac{1}{2} \sum_{j}(1-\mu_{j}^{z}).
\end{align}

The followed calculations are direct and similar to previous ones. We will omit the details. In short, we get the same expressions of  final excitation probability and density of defects as the ones in Eqs. (\ref{excitation_f}) and (\ref{simp-dd}) respectively. Moreover, the total dynamical phase $p_q^f$ and the period of oscillation $T_Q$ are the same as the ones in Eqs. (\ref{psi_gt}) and (\ref{period-g}), but note that we have $g_{\text{rt}}>g_c=1$ now. The oscillatory density of defects for $R=1$ and $g_{\text{rt}}=1.5$ is illustrated in Fig. \ref{density-reverse}.

There is a little difference for the amplitude of the oscillatory density of defects. If we fix the value of $g_{\text{rt}}~(>1)$, its asymptotical behavior still falls into $\lim_{\tau_{Q}\rightarrow\infty}M\sim(\ln\tau_Q)^{-3/2}$. But, if letting $g_{\text{rt}}\rightarrow\infty$, we get
\begin{align}
  \lim\limits_{g_{\text{rt}}\rightarrow\infty}M=\sqrt{R}\left[\frac{\pi}{(1+R)(g_{\text{rt}}-1)}\right]^{3/2}~\rightarrow~0,
\end{align}
which means the oscillation will fade out eventually as a dephasing effect.

\section{Quantum $XY$ chain}
\label{XYchain}

The round-trip quench protocol can also be realized in the quantum $XY$ chain. More interestingly, we shall introduce another typical scheme, the \emph{quarter-turn quench protocol}, that can produce the same interference effect.

\begin{figure}[t]
  \begin{center}
	\includegraphics[width=3.2in,angle=0]{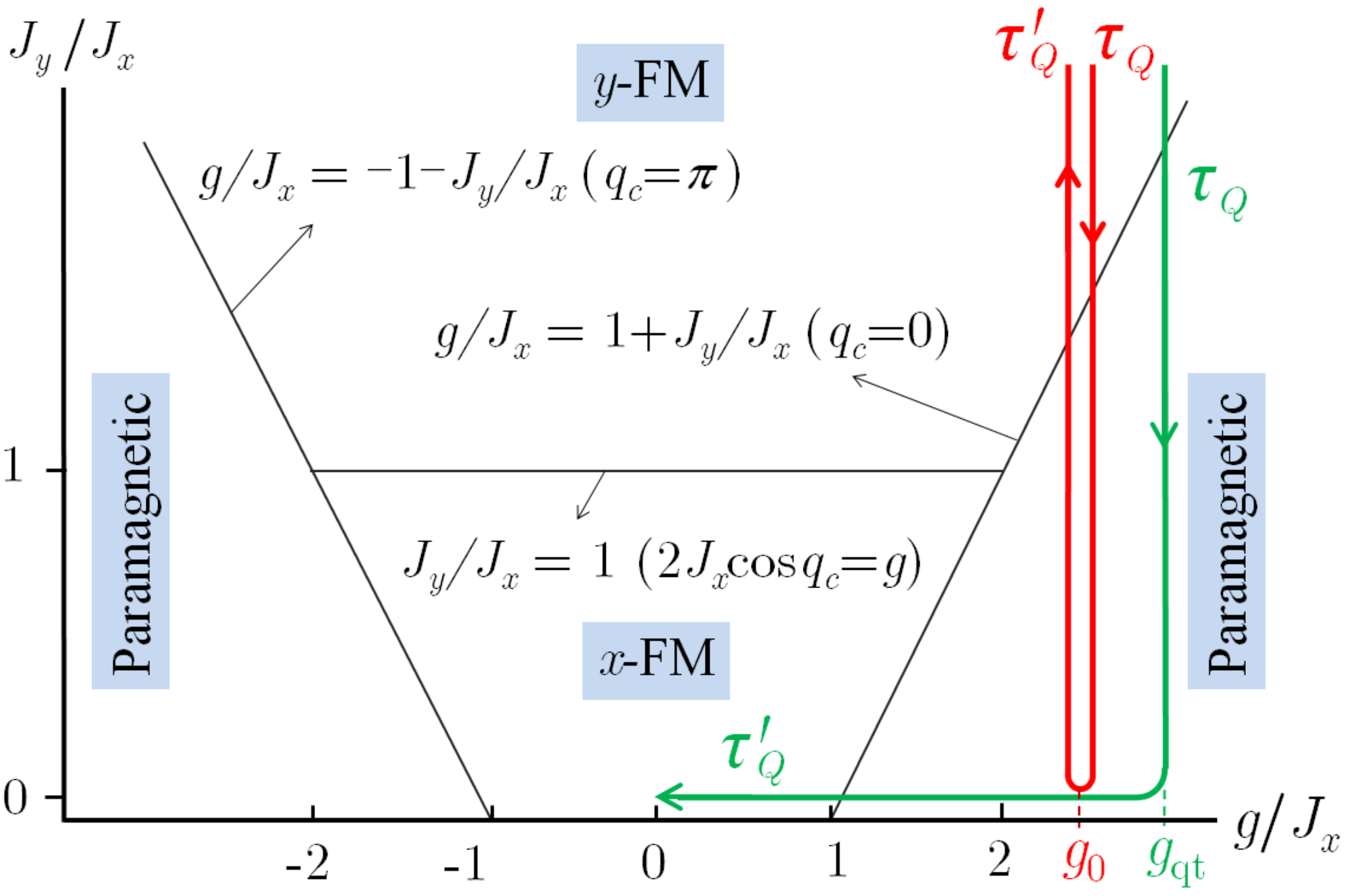}
  \end{center}
  \caption{Phase diagram of the quantum $XY$ model. Two quench protocols are illustrated by the colored lines, the round-trip (red) and quarter-turn ones (green). For the round-trip quench protocol, we mainly focus on the case that the system is driven across the tricritical point twice ($g_0=2$). In the quarter-turn quench, starting from the $y$-FM phase, the system goes across the phase boundary $g/J_{x}=1+J_{y}/J_{x}$ twice and finally reaches the limit of the $x$-FM phase, the classical Ising model. The turning point is labelled by the parameters fulfilling $g_{\text{qt}}>1$ and $J_{y}/J_{x}=0$. Please see more details in the text.}
  \label{XY-phase}
\end{figure}

The quantum $XY$ chain with a transverse field reads
\begin{equation} \label{XY-Hamiltonian}
	H_{XY}=-\sum_{j=1}^{N}\left( J_{x}\sigma_{j}^{x}\sigma_{j+1}^{x}+J_{y}\sigma_{j}^{y}\sigma_{j+1}^{y}+g\sigma_{j}^{z}\right),
\end{equation}
where $J_{y}~(>0)$ and $J_{x}~(>0)$ are interactions in $x$ and $y$ directions respectively. We shall set $J_{x}$ as an energy unit appropriately. The Hamiltonian can also be solved by Jordan-winger transformation \cite{Lieb_1961, Pfeuty_1970}. As shown in Fig. \ref{XY-phase}, this model exhibits four phases: two ferromagnetic (FM) and two paramagnetic phases. One gets $x$-FM or $y$-FM phase if $J_x$ or $J_y$ prevails respectively. In Fourier space, the quasiparticle dispersion reads
\begin{equation}
	\omega_{q}=2\sqrt{\left\lbrace g-(J_{x}+J_{y})\cos q\right\rbrace ^{2}+\left\lbrace (J_{x}-J_{y})\sin q\right\rbrace ^{2}}.
\end{equation}
On the phase boundaries, the gap between the ground state and the lowest excited state vanishes at a critical quasimomentum $q_c$. The phase boundaries are three lines: $g=J_{x}+J_{y}$, $g=-J_{x}-J_{y}$, and $J_{y}=J_{x}$ with $q_c=0$, $\pi$, and $\arccos(g/2J_{x})$ respectively. Moreover, there are two tricritical points at $(g/J_{x},~J_{y}/J_{x})=(\pm 2,~1)$.

\subsection{Round-trip quench protocol}

First of all, one can design several round-trip quench protocols on this system by defining appropriate defects in the final targeted state. However, they are not much different from the previous ones. Here we only remark on some new situations. Let us focus on the protocol that is represented by the red line along the ordinate $J_y/J_x$ axis ($g_0\geq 2$) in Fig. \ref{XY-phase}, in which starting and ending points are located in the deep region of the $y$-FM phase and the turning point is set at $J_{y}=0$. Without surprising, we have obtained the oscillatory density of defects that falls into the same formula as that expressed by Eq. (\ref{simp-dd}). However,  there emerges a new situation in which the system comes across the tricritical point ($g_{0}=2$) so we get revised factors,
\begin{equation}
  n_{0}=\left\lbrace
    \begin{array}{ccc}
      \frac{1}{2\pi(g_{0}-2)\sqrt{2\tau_{Q}}} & (g_{0}>2),\\
      \frac{\Gamma(7/6)}{\pi(2\pi\tau_{Q})^{1/6}}& (g_{0}=2),
\end{array}
    \right.
\end{equation}
and
\begin{equation}
  f = \left\lbrace
    \begin{array}{ccc}
      1+\frac{1}{\sqrt{R}}-\frac{2}{\sqrt{1+R}} & (g_{0}>2),\\
      1+\frac{1}{R^{1/6}}-\frac{2}{(1+R)^{1/6}} & (g_{0}=2).
    \end{array}
    \right.
\end{equation}
In both situations, the periods share the same expression,
\begin{equation}
  T_Q = \frac{\pi}{(g_0 -1)^2(1+R)}. \label{TQ_g0}
\end{equation}
This result indicates that the interference can occur not only between two critical dynamics but also between two tricritical dynamics. Our result coheres with the underlying QKZM for tricritical point revealed in previous study \cite{Sen_2007}.

\subsection{Quarter-turn quench protocol}

\begin{figure}[t]
  \begin{center}
    \includegraphics[width=3.2in,angle=0]{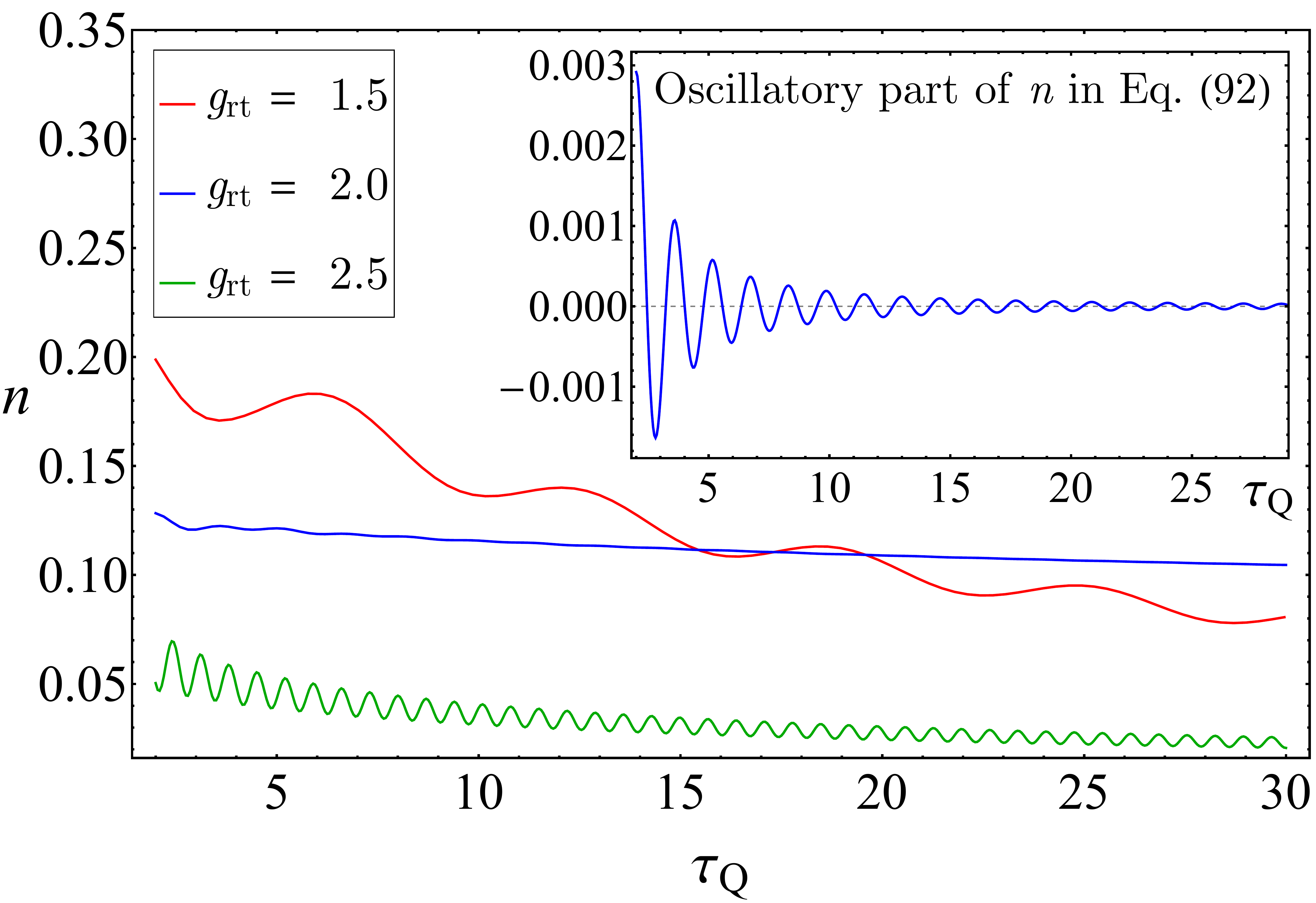}
  \end{center}
  \caption{Density of defects after the quarter-turn quench protocol is applied to the $XY$ chain with parameters $R=1$ and $g_{\text{qt}}=1.5~\text{(red)},~2~\text{(blue)},~2.5~\text{(green)}$. For the case of $g_{\text{qt}}=2$, the system will go across a tricritical point, thus the blue curve is determined by Eq. (\ref{n32}), whose oscillatory part is illustrated in the inset. The corresponding period of oscillation for the curves are $2\pi$, $\pi/2$, and $2\pi/9$ respectively in accordance with Eq. (\ref{TQ_XY}).}
  \label{density-XY}
\end{figure}

It is interesting to find new ways for realizing interference effect in the dynamics of this system, since it contains more abundant phases and phase transitions. We propose another typical case, the quarter-turn quench protocol, which is shown in Fig. \ref{density-XY}. The full procedure also contains two linear ramps and can be parameterized as
\begin{equation}
\left\lbrace \begin{array}{llc}
J_{y}(t)=-\frac{t}{\tau_{Q}}~\text{and}~ g(t)=g_{\text{qt}} ~&(t_{i}<t\leq 0),\\
J_{y}(t)=0~\text{and}~ g(t)=g_{\text{qt}}-\frac{t}{\tau'_{Q}} ~&(0<t\leq t_{f}),
\end{array}
\right.
\end{equation}
where $t_{i}=-\infty$ and $t_{f}=g_{\text{qt}}\tau'_{Q}$. The starting point is set in the deep region of the $y$-FM phase. In the first stage, the interaction $J_y$ is ramped down to zero so that the system is driven to the paramagnetic phase, which is ensued by the transverse field taking an appropriate value. Then, in the second stage, the transverse field is ramped down from $g_{\text{qt}}>1$ to zero and the system reaches the classical Ising model eventually. In this limit, again, the kinks along the $x$ axis play the role of defects and the number of kinks matches the number of excitations exactly according to Eq. (\ref{IsingKink-x}).

Because the situation is much more delicate now, we write down the final excitation probability before taking long wave approximation,
\begin{align}
	p_{q}^{f}=\left\lbrace
    \begin{array}{l}
    A^{2}+B^{2}-2AB\cos\psi~~~~~~(1<g_{\text{qt}}<2),\\
    A^{2}+B^{2}-2AB\cos(\psi-\pi)~~~~(g_{\text{qt}}\geq2),
    \end{array}
    \right. \label{pqf_XY}
\end{align}
where
\begin{align}
  A&=e^{-\pi\tau_{Q}(g_{\text{qt}}\sin q-\sin 2q)^{2}}\sqrt{1-e^{-2\pi \tau'_{Q} \sin^{2}q}},\\
  B&=e^{-\pi \tau'_{Q} \sin^{2}q}\sqrt{1-e^{-2\pi\tau_{Q} (g_{\text{qt}}\sin q-\sin 2q)^{2}}},\\
  \psi&=\frac{\pi}{2}+2(g_{\text{qt}}\cos q-\cos 2q)^{2}\tau_{Q}+2(g_{\text{qt}}-\cos q)^{2}\tau'_{Q} \nonumber\\
      &+\tau_{Q}(g_{\text{qt}}\sin q-\sin 2q)^{2}\left[ \ln\{4\tau_{Q}(g_{\text{qt}}\cos q-\cos 2q)^{2}\}\right.\nonumber\\
  &~+\left.\gamma_{E}\right] +\tau'_{Q}\sin^{2}q\left[ \ln\{4\tau'_{Q}(g_{\text{qt}}-\cos q)^{2}\}+\gamma_{E}\right]. \label{psi_qt}
\end{align}
The delicacy lies in the following facts.

First, the system undergoes two and three quantum phase transitions for $g_{\text{t}}>2$ and $1<g_{\text{t}}<2$ respectively. In the former case, the two transitions occur at the same phase boundary $g=J_{x}+J_{y}$ with critical quasimomentum $q_c = 0$, thus an oscillation in the density of defects would be observed inevitably. In the latter case, an extra transition occurs at the phase boundary $J_{y}/J_{x}=1$ with critical quasimomentum $q_c = \arccos(g/2J_{x})$, which is independent from the other two transitions and does not affect their interference. In both cases, the final density of defects still takes the general form in Eq. (\ref{simp-dd}) but with renewed factor
\begin{equation}
	f=\left\lbrace
	\begin{array}{ccc}
     \frac{1}{g_{\text{qt}}-2}+\frac{1}{\sqrt{R}}-\frac{2}{\sqrt{(g_{\text{qt}}-2)^{2}+R}} & (g_{\text{qt}}>2),\\
     \frac{6+g_{\text{qt}}}{4-g_{\text{qt}}^{2}}+\frac{1}{\sqrt{R}}-\frac{2}{\sqrt{(g_{\text{qt}}-2)^{2}+R}} & (1<g_{\text{qt}}<2),
	\end{array}
	\right.
\end{equation}
and period of oscillation,
\begin{align}
  T_{Q}=\frac{\pi}{(g_{\text{qt}}-1)^{2}(1+R)}. \label{TQ_XY}
\end{align}
We have also verified numerically that the amplitude of oscillation behaves asymptotically like $\lim_{\tau_{Q}\rightarrow\infty}M\sim(\ln\tau_Q)^{-3/2}$.

Second, if we let $g_{\text{qt}}=2$, the system will go across a tricritical point in the first linear ramp and a usual critical point in the second linear ramp. It is well-known that a single tricritical point will lead to a quite different scaling, $n\sim\tau_Q^{-1/6}$, rather than the familiar scaling, $n\sim\tau_Q^{-1/2}$, fit for the usual critical point. After the quarter-turn quench process, the final density of defects should contains both contributions intricately. Although no common QKZM factor (like $n_0$ in Eq. (\ref{n0})) could be singled out, the final density of defect would reflect an interplay between the critical point and tricritical point. Specifically, we arrive at
\begin{eqnarray}
  n=&\frac{\Gamma(7/6)}{\pi(2\pi\tau_{Q})^{1/6}}+\frac{1}{2\pi\sqrt{2\tau'_{Q}}}-\frac{\pi^{1/3}\left\lbrace \text{Ai}(x)^{2}+\text{Bi}(x)^{2}\right\rbrace }{\sqrt{2}(3\tau_{Q})^{1/6}}\nonumber\\
	&+A'\cos\left(2\pi\frac{\tau_{Q}}{T_{Q}}+\delta\right),\label{n32}
\end{eqnarray}
where $\text{Ai}(x)$ and $\text{Bi}(x)$ are two kinds Airy functions and $x=-\frac{\pi^{2/3}\tau'_{Q}}{(3\tau_{Q})^{1/3}}$. The first two terms are individual contributions from two linear ramps respectively. The third term is the nonoscillatory part of the interplay between the two ramps. The last term is the oscillatory part (Please see the inset in Fig. \ref{density-XY}), whose amplitude $A'$ goes to zero quickly when $\tau_Q$ is large enough with the asymptotical behavior, $\sim\tau_Q^{-3/2}$.

The final densities of defects for three distinct values of $g_{\text{qt}}$ are illustrated in Fig. \ref{density-XY}, in which the curves are obtained by integration on $p_q^f$ expressed in Eq. (\ref{pqf_XY}) over the first Brillouin zone. And we have verified that a numerical solution of the dynamical TDBdG equation gives almost the same results. We can see that the oscillations for $g_{\text{qt}}\neq2$ are prominent. While the oscillation for $g_{\text{qt}}=2$ fades out very quickly with $\tau_Q$ increasing, although the period is still given by Eq. (\ref{TQ_XY}).

\section{Defect-defect correlator with multiple length scales}
\label{transversecorrelator}

In this section, we disclose an interesting phenomenon of multiple length scales, diagonal and off-diagonal ones, in the defect-defect correlator due to the interference effect. This correlator can reflect the special dephasing effect in the post-transition state. By a comparative study of the round-trip and reversed round trip quench protocols for the transverse Ising chain, we show that the dephased result relies on how the diagonal and off-diagonal lengths are modulated by the controllable parameter in a quench protocol.

Throughout this paper, we only concern two kinds of definitions of defects according to the destination of dynamics: $\mathcal{N}_P$ in Eq. (\ref{definition-defect}) for the limit of paramagnetic phase and $\mathcal{N}_F$ in Eq. (\ref{IsingKink-x}) for the limit of ferromagnetic phase. The former is fit for the round-trip quench protocol, while the latter the reversed round-trip quench protocol. Likewise, the defect-defect correlator should be defined differently for these two limits. For the round-trip quench protocol, it turns out to be the transverse spin-spin correlator \cite{Das_2021},
\begin{align}
  C_{r}^{zz}&=\langle\hat{P}_j\hat{P}_{j+r}\rangle - \langle\hat{P}_{j}\rangle \langle\hat{P}_{j+r}\rangle\nonumber\\
            &=\frac{1}{4}(\langle\sigma_{j}^{z}\sigma_{j+r}^{z}\rangle - \langle\sigma_{j}^{z}\rangle \langle\sigma_{j+r}^{z} \rangle),
\end{align}
where the defect operator $\hat{P}_j=\frac{1}{2}(1-\sigma_{j}^{z})$. While for the reversed round-quench protocol, it is the kink-kink correlator \cite{Dziarmaga_2021},
\begin{align}
  C_{r}^{KK}&=\langle \hat{F}_j\hat{F}_{j+r} \rangle-\langle\hat{F}_{j}\rangle\langle\hat{F}_{j+r}\rangle
\end{align}
where the defect (or kink) operator $\hat{F}_j=\frac{1}{2}(1-\sigma_{j}^{x}\sigma_{j+1}^{x})$.

\subsection{Transverse spin-spin correlator for the round-trip quench protocol}

We consider the round-trip quench protocol applied to the transverse Ising chain first. For abbreviation, we confine our discussion to the simplest case with $R=1$ and $g_{\text{rt}}=0$.

\subsubsection{Fermionic correlators and multiple length scales}

The system goes back to the paramagnetic phase finally, thus the transverse correlator plays the role of defect-defect correlation. The details on computing this correlator are presented in Appendix \ref{Appendix_C}. Here we only quote the result:
\begin{equation}
	C_{r}^{zz}=|\beta_{r}|^{2}-\alpha_{r}^{2}, \label{Crzz}
\end{equation}
where
\begin{align}
  \alpha_{r}&\equiv\left\langle c_{j}^{+}c_{j+r}\right\rangle=-\int_{0}^{\pi}\frac{dq}{\pi}\left|v_{q}(t_{f})\right|^2 \cos(qr), \label{alpha-integration}\\
  \beta_{r}&\equiv\left\langle c_{j}c_{j+r}\right\rangle=\int_{0}^{\pi}\frac{dq}{\pi}u_{q}(t_{f})v_{q}^{*}(t_{f})\sin(qr),\label{beta-integration}
\end{align}
are diagonal and off-diagonal quadratic fermionic correlators \cite{Zurek_2007}. The negative term in Eq. (\ref{Crzz}) implies an antibunching effect of the defects in short space distances, which means the defects can hardly approach one another.

For the diagonal fermionic correlator $\alpha_{r}$, we arrive at
\begin{align}
  \alpha_{r}=
  &\frac{2e^{- r^2/\hat{\xi}^{2}}}{\sqrt{\pi}~\hat{\xi}}\left(1-\sqrt{2}e^{- r^2/\hat{\xi}^{2}}\right) \nonumber\\
  +&\sum_{m=1,2}\frac{\sqrt{8}(-1)^{m} e^{- r^2/\left(l^{\alpha}_{m}\right)^{2}}}{(2m\pi^{2})^{1/4}\sqrt{\hat{\xi}~l^{\alpha}_{m}}} \sin\phi_{m,r}^{\alpha},
  \label{alpha-r}
\end{align}
where
\begin{equation}
  \phi_{m,r}^{\alpha}= 4\tau_{Q}-\frac{b}{m}\frac{r^{2}}{(l_{m}^{\alpha})^{2}}
  +\frac{1}{2}\arctan\frac{b}{m} \label{phi_alpha}
\end{equation}
are phase factors with $b=\frac{1}{\pi}\left\{\ln(4\tau_{Q})+\gamma_{E}-2 \right\}\approx\frac{\ln\tau_Q}{\pi}$ and
\begin{align}
  &\hat{\xi}=4\sqrt{\pi\tau_{Q}}, \label{xi}\\
  &l'^{\alpha}_{m}=2\sqrt{2m\pi\tau_{Q}}\sqrt{1+\left(\frac{b}{m} \right)^{2} }, \label{l_alpha}
\end{align}
are three length scales. Exposed by the interference, $l_m^{\alpha}$'s ($m=1,2$) are two new lengths that could be called the \emph{diagonal lengths} since they appear in the diagonal fermionic correlator. They attenuate the sinusoidal interference terms, $\sin\phi_{m,r}^{\alpha}$'s, by the Gaussian decaying factors in space. We still call $\hat{\xi}$ the \emph{KZ length}, although it also appears in the diagonal fermionic correlator.

\begin{figure}[t]
  \begin{center}
	\includegraphics[width=3.2in,angle=0]{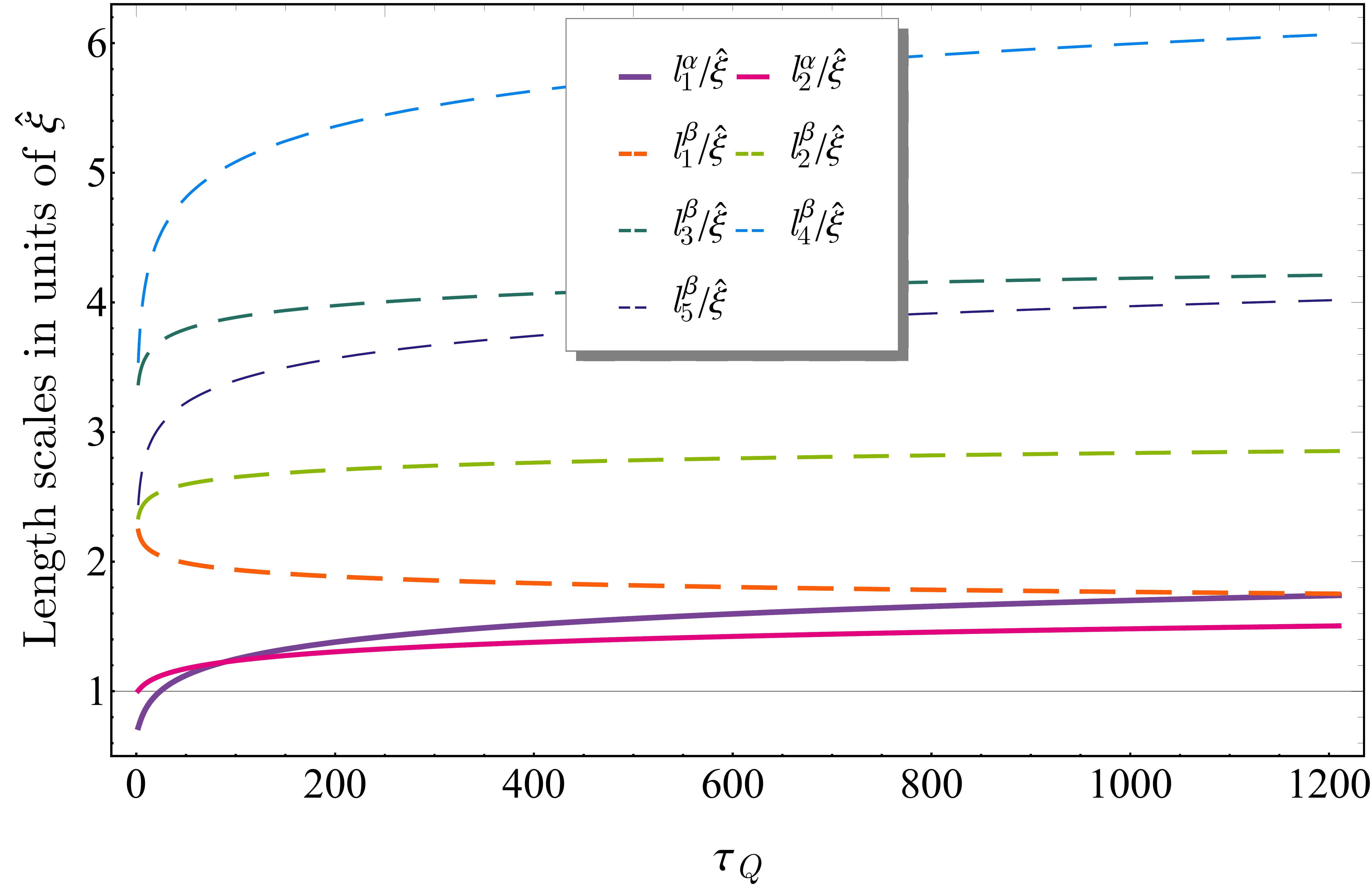}
  \end{center}
  \caption{Length scales, $l_{m}^{\alpha}(m=1,2)$ and $l_{m}^{\beta} (m=1,...,5)$, in the transverse spin-spin correlator. All length scales are plotted in units of KZ length $\hat{\xi}$. We have fixed the parameters, $R=1$ and $g_{f}=10$. }
  \label{length-scale}
\end{figure}

\begin{figure}[t]
  \begin{center}
	\includegraphics[width=3.0in,angle=0]{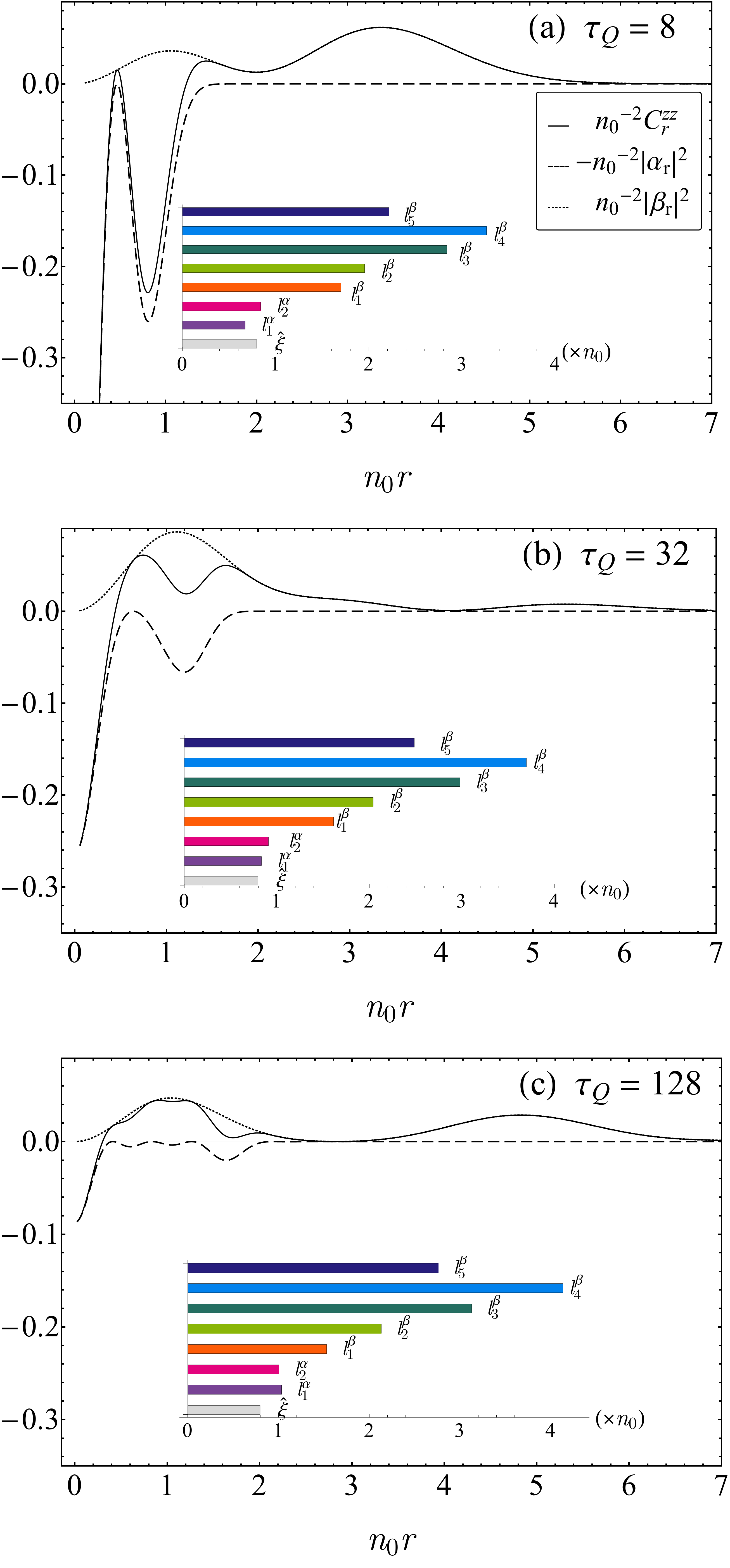}
  \end{center}
  \caption{Transverse spin-spin and fermionic correlators. The scaled correlators, $C_{r}^{zz}$, $-|\alpha_{r}|^{2}$, and $|\beta_{r}|^{2}$, are plotted in function of scaled distance $n_0 r$ for (a) $\tau_Q=8$, (b) $\tau_Q=32$, and (c) $\tau_Q=128$. Other parameters are $R=1$ and $g_{f}=10$. The multiple lengths are illustrated by the colored bar charts in each insets. The fermionic correlator, $-\alpha_{r}^{2}$, also stands for the dephased transverse spin-spin correlator when $g_f\rightarrow\infty$ according to Eq. (\ref{dephasedC}).}
  \label{transverse-correlation}
\end{figure}

For the off-diagonal fermionic correlator $\beta_{r}$,  we arrive at
\begin{equation}
  \beta_{r}=\sum_{m=1}^{5}(-1)^{m-1}\frac{y_{m}~r}{\sqrt{\hat{\xi}~(l_{m}^{\beta})^{3}}} ~e^{-r^2/\left(l_{m}^{\beta}\right)^{2} }e^{i \phi_{m}^{\beta}(r)},\label{beta-r}
\end{equation}
where $y_{m}=\frac{2Y_{m}}{\left\{\pi^2(h_{m})^{3}\right\}^{1/4}}$ are coefficients with $Y_{2}=\sqrt{\frac{e}{\ln2}}$ for $m=2$ and $Y_{m}=\frac{1}{2}\sqrt{\frac{e}{\ln2}}$ for other $m$,
\begin{equation}
  \phi_{m}^{\beta}(r)=\lambda'_{m}-\frac{\lambda_{m}~r^{2}}{\pi h_{m}(l_{m}^{\beta})^{2}}-\frac{3}{2}\arg\left( 1-i\frac{\lambda_{m}}{\pi h_{m}}\right) \label{phi_beta}
\end{equation}
are phase factors with
\begin{equation}
  \left.
  \begin{array}{ll}
    &\lambda_{1}=\ln\tau_{Q}-2g_{f}-2\ln(g_{f}-1),\\
    &\lambda_{2}=\lambda_{3}=-\ln\tau_{Q}-2g_{f}-2\ln(g_{f}-1),\\
	&\lambda_{4}=\lambda_{5}=-3\ln\tau_{Q}-2g_{f}-2\ln(g_{f}-1),\\
    &\lambda'_{1}=-\lambda'_{2}=-\lambda'_{3}=\frac{\pi}{4}+2\tau_{Q},\\
    &\lambda'_{4}=\lambda'_{5}=-\frac{3\pi}{4}-6\tau_{Q},\\
    &h_{1}=h_{2}=h_{5}=2+\frac{1}{\ln2},~h_{3}=h_{4}=\frac{1}{\ln2},
  \end{array}
  \right.\nonumber
\end{equation}
and
\begin{equation}
  l_{m}^{\beta}=2\sqrt{\pi h_{m}\tau_{Q}}\sqrt{1+\left(\frac{\lambda_{m}}{\pi h_{m}} \right)^{2} } \label{l_beta}
\end{equation}
are other five new lengths. Likewise, we call them \emph{off-diagonal lengths}.

So we get eight lengths in total. We illustrate them in Fig. \ref{length-scale} by setting the typical parameters $R=1$ and $g_{f}=10$. It is clear to see that $l_m^{\beta}$'s are larger than $l_m^{\alpha}$'s and $l_{4}^{\beta}$ is always the largest one overall. In fact, except for the KZ length ($\hat{\xi}\sim\sqrt{\tau_Q}$), all other lengths share the same asymptotic behavior, $\sim\sqrt{\tau_Q}\ln\tau_Q$, for a fixed finite value of $g_f$.

The transverse spin-spin correlator scaled as $n_0^{-2}C_r^{zz}$ versus the scaled distance $n_0 r$ is exemplified in Fig. \ref{transverse-correlation}. We see that the length scales coming out of the diagonal and off-diagonal fermionic correlators play quite different roles. The transverse correlator is governed by the diagonal part for small space distances and by the off-diagonal part for large space distances, i.e., we have
\begin{align}
	C_{r}^{zz}\approx \left\{
\begin{array}{cll}
-\alpha_{r}^{2}&~~(r \ll \min(l_{m}^{\alpha},\hat{\xi})),\\
|\beta_{r}|^{2}&~~(r \gtrsim \min(l_{m}^{\beta})),
\end{array}
    \right. \label{Crzzapp}
\end{align}
approximately. The former is due to the fact: $(l_m^{\alpha})^{-1/2}\gg r(l_m^{\beta})^{-3/2}$ for small $r$, and the latter: $e^{-(l_m^{\beta}/l_m^{\alpha})^2}\rightarrow 0$ and $e^{-(l_m^{\beta}/\hat{\xi})^2}\rightarrow 0$ for large enough $r$. When the space distance exceeds the largest length scale, i.e. $r > l_{4}^{\beta}$, the transverse correlator reduces to the Maxwell-Boltzmann form
\begin{equation}
  C_{r}^{zz}\approx\frac{1}{16}\sqrt{\frac{e^{2}\ln2}{\pi}}\frac{r^{2}}{\hat{\xi}~(l_{4}^{\beta})^{3}} ~e^{-2r^2/\left(l_{4}^{\beta}\right)^{2}}
\end{equation}
with only one prevailing length, $l_{4}^{\beta}$. While for intermediate space distances, the transverse correlator is influenced by all lengths.

\subsubsection{Periodicity due to interference}

\begin{figure}[t]
\begin{center}
  	\includegraphics[width=3.4in,angle=0]{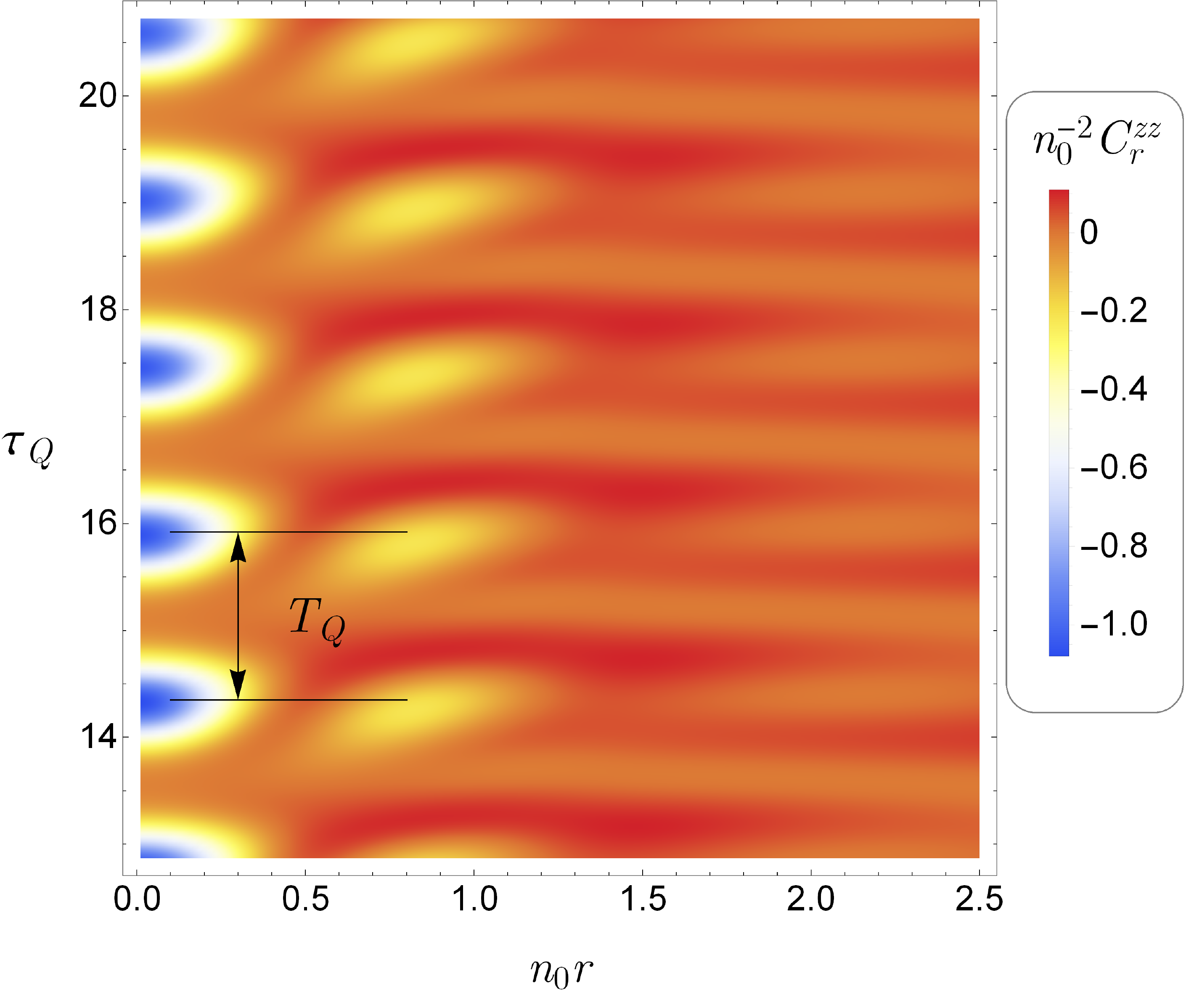}
  \end{center}
  \caption{Density plot of $n_0^{-2}C_r^{zz}$ in typical ranges of variables $n_0 r$ and $\tau_Q$. We have set the parameters $R=1$ and $g_{f}=10$. $T_Q=\frac{\pi}{2}$ denotes the period of the oscillation along the direction of the variable $\tau_Q$ in the transverse spin-spin correlator.}
  \label{densityCzzr}
\end{figure}

Both fermionic correlators, $\alpha_r$ and $\beta_{r}$, contain interference terms. It is easy to discern an oscillation with a period, $T_Q=\frac{\pi}{2}$ (Please note that $R=1$ at present), in the diagonal part of the transverse spin-spin correlator, $-|\alpha_r|^2$, along the $\tau_Q$ direction. While for the off-diagonal part, $|\beta_{r}|^2$, one can find that
\begin{align}
  |\beta_{r}|^2\sim\sum_{m,n=1,..5}A_{mn}\cos(\Omega_{mn}\tau_{Q}+\delta_{mn}),
\end{align}
where
\begin{align}
  \Omega_{mn}&\equiv\frac{\lambda'_{m}-\lambda'_{n}}{\tau_{Q}}-\frac{r^{2}}{\tau_{Q}^{2}}\left(\frac{1}{\lambda_{m}}-\frac{1}{\lambda_{n}} \right) \nonumber\\
  -&\frac{3}{2\tau_{Q}}\left[\arg\left( 1-i\frac{\lambda_{m}}{\pi h_{m}}\right)-\arg\left( 1-i\frac{\lambda_{n}}{\pi h_{n}}\right) \right].
\end{align}
$\Omega_{mn}$ takes two possible values, $4$ or $8$, asymptotically when $\tau_{Q}\gg r$ (Please note that we also have $\tau_{Q}\gg \hat{\xi}, l_{m}^{\beta}$). Thus the fermionic correlator $|\beta_{r}|^2$ also exhibits the same period, $T_Q=\frac{\pi}{2}$. The oscillatory behavior characterized by the period $T_Q$ can be easily observed in the density plot of the full transverse spin-spin correlator as shown Fig. \ref{densityCzzr}.

\subsubsection{Quantum dephasing}

The off-diagonal lengths, $l_{m}^{\beta}$'s, rely on the controllable parameter $g_{f}$. If we let $g_{f}\gg \ln \tau_{Q}$, instead of fixing $g_f$, all off-diagonal lengths will grow linearly with the final time $t_{f}=g_f \tau_Q$ increasing,
\begin{equation}
	l_{m}^{\beta}\sim \sqrt{\tau_{Q}}g_{f}=\frac{t_{f}}{\sqrt{\tau_{Q}}}.\label{lmgf}
\end{equation}
The off-diagonal fermionic correlator can be written as
\begin{align}
\beta_{r}=&\int_{0}^{\pi}\frac{dq}{\pi}\sin(qr)e^{-h_{m}\pi\tau_{Q}q^{2}}X_{m}\sqrt{2\pi\tau_{Q}}\nonumber\\
&\times e^{i(4t_{f}-2\frac{t_{f}^{2}}{\tau_{Q}}+2q^{2}t_{f})}\left(\sum_{m=1}^{5}(-1)^{m-1}e^{i\lambda'_{m}}\right).
\end{align}
We see the off-diagonal lengths, $l_m^{\beta}$'s, play the role of dephasing length, because the phase factor, $\exp\{i(4t_{f}-2\frac{t_{f}^{2}}{\tau_{Q}}+2q^{2}t_{f})\}$, oscillates very rapidly with the quasimomentum $q$ varying and the magnitude of $\beta_{r}$ becomes negligible when $t_{f}\rightarrow\infty$. The dephasing time $t_D$ measures the time when the dephasing effect becomes noticeable \cite{Dziarmaga_2021}. Here, it can be estimated by setting
\begin{align}
  \frac{\lambda_m}{\pi h_m} = 1
\end{align}
in Eq. (\ref{l_beta}). Now that we have leading terms linear in $g_f$, i.e. $|\lambda_m|\propto 2g_f = 2\frac{t_f}{\tau_Q}$, and there are two values of $h_m$ (i.e. $h_{1}=h_{2}=h_{5}=2+\frac{1}{\ln2}$ and $h_{3}=h_{4}=\frac{1}{\ln2}$), we get two dephasing times due to the multiple length scales,
\begin{align}
  &t_D^{3,4} = \frac{1}{2\ln2}\pi\tau_Q,\nonumber\\
  &t_D^{1,2,5} = \left(1+\frac{1}{2\ln2}\right)\pi\tau_Q,
\end{align}
which means that the off-diagonal fermionic correlator decreases significantly at two moments with $t_f$ increasing. Meanwhile, the KZ length and the two diagonal lengths remains intact after such a dephasing, which leads to a reduced transverse spin-spin correlator,
\begin{equation}
  C_{r}^{zz}(t_{f}\rightarrow\infty)\approx-\alpha_{r}^{2}. \label{dephasedC}
\end{equation}
This result is illustrated by the dashed lines in Fig. \ref{transverse-correlation}. Besides the strong antibunching, the dashed lines also display a sinusoidal behavior that is rendered by the diagonal lengths. The sinusoidal behavior is in contrast to the traditional case without interference, where only the KZ length remains \cite{Dziarmaga_2021}. Moreover, the periodicity in $\tau_Q$ direction still remains in the dephased correlator.

\subsection{Kink-kink correlator for the reversed round-trip quench protocol}

Now we point out that the same phenomenon of multiple length scales will also appear in the kink-kink correlator in the reversed round-trip quench protocol applied to the transverse Ising chain. But there are some interesting differences that need to be addressed adequately. We mainly focus on the dephasing effect that is regulated by the multiple lengths.

The Kink-Kink correlator after the reverse round-trip quench process can still be reduced to
\begin{equation}
C_{r}^{KK}=|\beta'_r|^2-(\alpha'_r)^2,
\end{equation}
where $\alpha'_r$ and $\beta'_r$ are diagonal and off-diagonal quadratic fermionic correlators.

For the diagonal fermionic correlator $\alpha'_{r}$, we arrive at
\begin{align}
\alpha'_{r}=
&\frac{2e^{- r^2/\hat{\xi}^{2}}}{\sqrt{\pi}~\hat{\xi}}\left(1-\sqrt{2}e^{- r^2/\hat{\xi}^{2}}\right) \nonumber\\
+&\sum_{m=1,2}\frac{\sqrt{8}(-1)^{m} e^{- r^2/\left(l'^{\alpha}_{m}\right)^{2}}}{(2m\pi^{2})^{1/4}\sqrt{\hat{\xi}~l'^{\alpha}_{m}}} \sin\phi_{m,r}^{'\alpha},
\end{align}
where
\begin{equation}
\phi_{m,r}^{'\alpha}= 4\tau_{Q}-\frac{b'}{m}\frac{r^{2}}{(l'^{\alpha}_{m})^{2}}
+\frac{1}{2}\arctan\frac{b'}{m}
\end{equation}
are phase factors with
\begin{align}
  b'=&2\left[ \ln\left\{4\tau_{Q}(g_{\text{rt}}-1)^{2}\right\}+2(g_{\text{rt}}-1)+\gamma_{E}\right],\\
  l'^{\alpha}_{m}=&\frac{1}{\pi}\sqrt{2m\pi\tau_{Q}}\sqrt{1+\left(\frac{b'}{m} \right)^{2} }, (m=1,2)
\end{align}
and $\hat{\xi}$ defined in Eq. (\ref{xi}) is the usual KZ length. Notably, the two lenght scales, $l'^{\alpha}_m$ ($m=1,2$), are difference from the ones of the transverse correlator in the round-trip quench protocol.

\begin{figure}[t]
  \begin{center}
	\includegraphics[width=3.3in,angle=0]{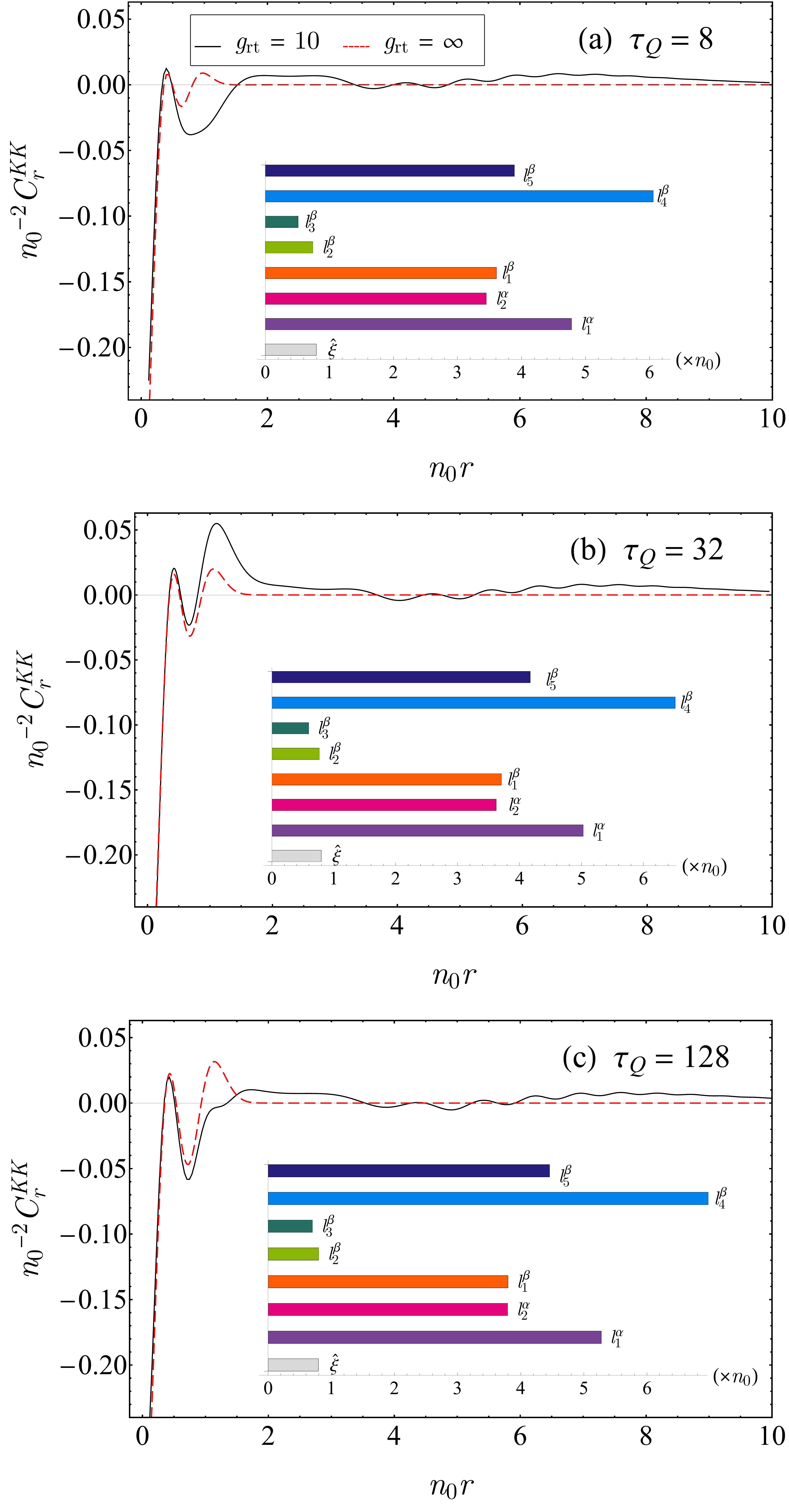}
  \end{center}
  \caption{Kink-kink correlator. The black solid lines and the colored bar charts of multiple lengths in the insets are for the case of $g_{\text{rt}}=10$. The red dashed lines are dephased results for $g_{\text{rt}}\rightarrow\infty$.}
  \label{Kink-correlation}
\end{figure}

For the off-diagonal fermionic correlator $\beta'_{r}$,  we arrive at
\begin{equation}
\beta'_{r}=\sum_{m=1}^{5}(-1)^{m-1}\frac{y_{m}~r}{\sqrt{\hat{\xi}~(l'^{\beta}_{m})^{3}}} ~e^{-r^2/\left(l'^{\beta}_{m}\right)^{2} }e^{i \phi_{m}^{'\beta}(r)},
\end{equation}
$y_{m}=\frac{2Y_{m}}{\left[\pi(h_{m})^{3}\right]^{1/4}}$ are coefficients with $Y_{2}=\sqrt{\frac{e}{\ln2}}$ for $m=2$ and $Y_{m}=\frac{1}{2}\sqrt{\frac{e}{\ln2}}$ for other values of $m$,
\begin{equation}
\phi_{m}^{'\beta}(r)=\chi'_{m}-\frac{\chi_{m}~r^{2}}{\pi h_{m}(l_{m}^{\beta})^{2}}-\frac{3}{2}\arg\left( 1-i\frac{\chi_{m}}{\pi h_{m}}\right)
\end{equation}
are phase factors with
\begin{equation}
\left.
\begin{array}{ll}
&\chi_{1}=\ln 4\tau_{Q}+4g_{\text{rt}}-2+4\ln(g_{\text{rt}}-1)+\gamma_{E},\\
&\chi_{2/3}=-(\ln 4\tau_{Q}-2+\gamma_{E}),\\
&\chi_{4/5}=-\{3\ln 4\tau_{Q}+4g_{\text{rt}}-6+4\ln(g_{\text{rt}}-1)+3\gamma_{E}\},\\
&\chi'_{1}=\frac{\pi}{4}+2\tau_{Q}(2g_{\text{rt}}^{2}-4g_{\text{rt}}+1),\\
&\chi'_{2/3}=-\frac{\pi}{4}-2\tau_{Q},\\
&\chi'_{4/5}=-\frac{3\pi}{4}-2\tau_{Q}(2g_{\text{rt}}^{2}-4g_{\text{rt}}+3),\\
&h_{1}=h_{2}=h_{5}=2+\frac{1}{\ln 2},~ h_{3}=h_{4}=\frac{1}{\ln 2},
\end{array}
\right.\nonumber
\end{equation}
and
\begin{equation}
l'^{\beta}_{m}=2\sqrt{\pi h_{m}\tau_{Q}}\sqrt{1+\left(\frac{\chi_{m}}{\pi h_{m}} \right)^{2} }
\end{equation}
are five more new length scales.

The dephasing process here is quite different from the former of the transverse spin-spin correlator. Now we notice the fact that the KZ length, $\hat{\xi}$, and the off-diagonal lengths, $l_{2}^{\beta}$ and $l_{3}^{\beta}$, are free of  $g_{\text{rt}}$, while all others increase linearly in $g_{\text{rt}}$: $l'^{\alpha}_{1/2},l'^{\beta}_{1/4/5}\sim \sqrt{\tau_{Q}}g_{\text{rt}}$. This fact means that only $\hat{\xi}$, $l_{2}^{\beta}$, and $l_{3}^{\beta}$ can remain intact with the parameter $g_{\text{rt}}\rightarrow\infty$. So, after such a dephasing, we get a reduced kink-kink correlator,
\begin{eqnarray}
	C^{KK}_{r}(g_{\text{rt}}\rightarrow\infty)=- \frac{4e^{- 2r^2/\hat{\xi}^{2}}}{\pi~\hat{\xi}^{2}}\left(1-\sqrt{2}e^{- r^2/\hat{\xi}^{2}}\right)^{2}\nonumber\\
	+\left| \sum_{m=2,3}^{}\frac{(-1)^{m-1}y_{m}~r~e^{-r^2/\left(l'^{\beta}_{m}\right)^{2} }e^{i \phi_{m}^{'\beta}(r)}}{\sqrt{\hat{\xi}~(l'^{\beta}_{m})^{3}}} \right| ^{2}.
\end{eqnarray}
This dephased kink-kink correlators for $\tau_Q=8,~32,~128$ are illustrated by the red dashed lines in Fig. \ref{Kink-correlation}, which are plotted in comparison with the kink-kink correlators with finite value of $g_{\text{rt}}=10$ (black lines). We can still observe the strong antibunching in short space distances. But the sinusoidal behavior is rendered by the off-diagonal lengths instead of the diagonal ones. More interestingly, despite a bit of oscillatory behavior, the dephased kink-kink correlator can even become positive due to the contributions of the two off-diagonal lengths, $l_{2}^{\beta}$ and $l_{3}^{\beta}$.

\section{Summary and discussion}
\label{summary}

In summary, we have revealed a kind of interference effect by imposing appropriately designed quench protocol on the transverse Ising and quantum $XY$ models. The underlying mechanism can be well described by the combination of two successive Landau-Zener transitions, which renders an exposure of the dynamical phase in the final excitation probability so that the characteristic lengths in it are embedded in the density of defects.

In essence, the density of defects can reflect  the interplay between two same or different critical dynamics in a neat way. Let us retrospect a typical one as shown in Eq. (\ref{n32}): its first two terms are individual contributions of the two critical dynamics and the other two terms are joint contributions, in which one is nonoscillatory and the other oscillatory. On the practical side, the interference effect can be directly observed by the coherent many-body oscillation in the density of defects with predictable period $T_Q$. The formulae of the period that we discovered for several typical quench protocols suggest a quite generic form (e.g. please see Eqs. (\ref{period-g}), (\ref{TQ_g0}), and (\ref{TQ_XY})). But whether the formulae can represent a universal one or just a small sample needs more studies in diverse systems. This period comes from the term proportional to $\tau_Q+\tau'_Q = (1+R)\tau_{Q}$ in the dynamical phase (e.g. Eqs. (\ref{app-psi}), (\ref{psi_gt}), and (\ref{psi_qt})). In the dynamical phase, another term proportional to $(q-q_c)^{2}$ also plays important roles. First, it affects the amplitude of oscillation in the density of defects. Second, it can result in the phenomenon of multiple length scales in the defect-defect correlator and the dephased result relies on how the diagonal and off-diagonal lengths are modulated by the controllable parameter in a quench protocol.

In view of the current technological advances, e.g. the experiment in which a transverse Ising chain was emulated by Rydberg atoms \cite{Sachdev_2019}, or other ones that pave the way to study quantum dynamics \cite{Ebadi_2021, Scholl_2021, Semeghini_2021, Satzinger_2021}, the interferometry proposed in this paper provides a prospect method that may be used as a diagnostic tool to benchmark the experimental implementations of emulation against loss of coherence or simulators in favor of quantum computing.

\section*{ACKNOWLEDGMENTS}
We thanks Adolfo del Campo for pointing out the similarity of our round-trip quench protocol and the quench echo protocol that was considered in a previous work in Ref. \cite{Quan_2010} in the context of quantum adiabaticity. We also thanks Yan He and Shihao Bi for useful discussion. This work is supported by NSFC under Grants No. 1107417.

\appendix

\section{Excitation probability and final Density of defects} \label{Appendix_A}

By Eq. (\ref{Exci}), we can deduce the final excitation probability as
\begin{align}
  p_{q}=&\left|v_{q}(t_{f})u_{q}^{f}-u_{q}(t_{f})v_{q}^{f}\right| ^{2}\nonumber\\
  =&\left|-A~e^{i(\theta_{q}^{u}+\phi_{q}^{a})}+B~e^{-i(\theta_{q}^{v}+\phi_{q}^{b})}\right|^{2}\nonumber\\
  =& A^{2}+B^{2}-2AB\cos\psi,
\end{align}
where $u_{q}^{f}\approx1$ and $v_{q}^{f}\approx 0$ are the equilibrium Bogoliubov amplitudes defined in Eq. (\ref{Bogo}) at $g=g_{f}\gg 1$. $A$, $B$ and $\psi$ are defined in Eqs. (\ref{equation_A}), (\ref{equation_B}), and (\ref{equation_psi}) in the main text.

The density of defects is defined in Eq. (\ref{dod}). In the thermodynamical limit $N\rightarrow\infty$, we can replace the sum with an integral,
\begin{equation}
  n=\int_{0}^{\pi}\frac{\mathrm{d} q}{\pi}p_{q}.
\end{equation}
First, we make an approximation,
\begin{align}
  &\sqrt{(1-e^{-2\pi\tau_{Q}q^{2}})(1-e^{-2\pi R\tau_{Q}q^{2}})}\nonumber\\ \approx&\frac{1}{2}\left[\sqrt{R}(1-e^{-2\pi\tau_{Q}q^{2}})
  +\sqrt{\frac{1}{R}} (1-e^{-2\pi R \tau_{Q}q^{2}})\right],
\end{align}
Then, the defect density can be worked out by the integral formula,
\begin{equation}
\int_{0}^{\pi}\frac{\mathrm{d} q}{\pi}e^{-cq^{2}}\cos( a+bq^{2})=\frac{\cos\left[ a+\frac{\pi}{4}-\frac{1}{2}\arctan(c/b)\right] }{2\sqrt{\pi}(b^{2}+c^{2})^{1/4}},
\end{equation}
and the solution is
\begin{align}
  n = n_0 \left\{f + \sum_{i=1}^{3} M_{i}\cos(\Omega_{Q} \tau_Q+\delta_{i})\right\}, \label{n}
\end{align}
where
\begin{eqnarray}
  &f =1+\frac{1}{\sqrt{R}}-\frac{2}{\sqrt{1+R}},\\
  & \Omega_{Q}=2(1+R),\\
  &M_{i} = \frac{X_{i}}{c_{i}}\left\{1+\left(\frac{b}{c_{i}}\right)^{2}\right\}^{-1/4},\\
  &b=\frac{1}{\pi}\left[(1+R)\{\ln(4\tau_{Q})+\gamma_{E}-2\}+R\ln R\right], \label{b}\\
  &\delta_{i} = \frac{3\pi}{4}-\frac{1}{2}\arctan\frac{c_{i}}{b},\\
  &X_{1} = -\frac{2(1+R)}{\sqrt{2R}},~X_{2} = \sqrt{2R},~X_{3}=\sqrt{\frac{2}{R}},\\
  & c_{1}=1+R,~c_{2}=3+R,~c_{3}=1+3R.
\end{eqnarray}
We can rewrite the solution in Eq. (\ref{n}) to
\begin{align}
  n = n_0 \left\{f + M \cos(\Omega_{Q} \tau_Q+\delta)\right\}, \label{n2}
\end{align}
where the amplitude and phase are
\begin{eqnarray}
  &M = \sqrt{(\sum_{i=1}^{3}M_i\sin\delta_i)^2 + (\sum_{i=1}^{3}M_i\cos\delta_i)^2},\\
  &\delta =\arctan\frac{\sum_{i=1}^{3}M_i\sin\delta_i}{\sum_{i=1}^{3}M_i\cos\delta_i}.
\end{eqnarray}

\section{Round-trip quench protocol with the turning point, $g_{\text{rt}}=g_{c}=1$}
\label{turningpoint}
In the round-trip quench protocol with $g_{\text{rt}}=g_{c}=1$, we need to tackle the TDBdG equations in Eq.(\ref{time-BDG}) in another way. At the end of the first stage, $t=0$, the solutions are
\begin{align}
	v_{q}^{0}&=\frac{C_{1}\sqrt{\pi/2}}{\Gamma\left(1-i\tau_{Q}q^{2}/2\right)}2^{i\tau_{Q}q^{2}/2},\\
	u_{q}^{0}&=\frac{(-1)^{1/4}C_{1}\sqrt{\pi}}{\sqrt{\tau_{Q}}q~\Gamma\left(1/2-i\tau_{Q}q^{2}/2\right)}2^{i\tau_{Q}q^{2}/2},
\end{align}
where $C_{1}$ is defined in Eq. (\ref{C1C2}) and $2^{i\tau_{Q}q^{2}/2}$ is a trivial term. And, at the end of the second stage, the solutions are
\begin{align}
  v_{q}^{f}=&\frac{(-1)^{1/4}u_{q}^{0}}{\sqrt{2\pi\tau'_{Q}}~q}\left(e^{-\frac{3}{4}\pi\tau'_{Q}q^{2}}-e^{\frac{1}{4}\pi\tau'_{Q}q^{2}}\right)\Gamma\left(1-\frac{i}{2}\tau'_{Q}q^{2}\right)\nonumber\\
  &+\frac{v_{q}^{0}}{2\sqrt{\pi}}\left(e^{-\frac{3}{4}\pi\tau'_{Q}q^{2}}+e^{\frac{1}{4}\pi\tau'_{Q}q^{2}}\right)\Gamma\left(\frac{1}{2}-\frac{i}{2}\tau'_{Q}q^{2}\right),\\
  u_{q}^{f}=&\left\lbrace \frac{\sqrt{\tau'_{Q}}q~ v_{q}^{0}}{2\sqrt{\pi}(-1)^{1/4}}\frac{\Gamma\left(\frac{1}{2}-\frac{i}{2}i\tau'_{Q}q^{2}\right)}{\Gamma\left(1-\frac{i}{2}\tau'_{Q}q^{2}\right)}+\frac{u_{q}^{0}}{\sqrt{2\pi}}\right\rbrace  \nonumber\\ &\times\frac{\sqrt{2\pi}~\Gamma\left(1-\frac{i}{2}\tau'_{Q}q^{2}\right)}{\Gamma\left(1-i\tau'_{Q}q^{2}\right)}e^{-\frac{\pi}{4}\tau'_{Q}q^{2}}.
\end{align}
The final excitation probability is still expressed by Eq. (\ref{excitation_f}) but with different variables,
\begin{align}
  A=&\frac{1}{2}\sqrt{1+e^{-\pi\tau_{Q}q^{2}}}\sqrt{1-e^{-\pi\tau'_{Q}q^{2}}},\nonumber\\
  B=&\frac{1}{2}\sqrt{1-e^{-\pi\tau_{Q}q^{2}}}\sqrt{1+e^{-\pi\tau'_{Q}q^{2}}},\nonumber
\end{align}
and $\psi$ that has been shown in Eq. (\ref{exci-critical}). When $R=1$, we can get a reduced excitation probability,
\begin{equation}
  p_{q}^{f}=\left(1-e^{-2\pi\tau_{Q}q^{2}}\right)\sin^{2}\left[\frac{\pi}{4}+\frac{\tau_{Q}q^{2}}{2}\left\lbrace\gamma_{E}+\gamma_{d}\left(\frac{1}{2}\right)\right\rbrace \right] .
\end{equation}

\section{Calculation of the diagonal and off-diagonal fermionic correlators}\label{Appendix_C}

By the long wave approximation, we can rewrite the diagonal fermionic correlator in Eq. (\ref{alpha-integration}) to
\begin{align}
	\alpha_r=&\int_{0}^{\pi}\frac{dq}{\pi}\cos(qr)\left\lbrace
	2e^{-2\pi\tau_{Q}q^{2}}(e^{-2\pi\tau_{Q}q^{2}}-1)
	 \right.\nonumber\\
	&\left. +2(e^{-2\pi\tau_{Q}q^{2}}-e^{-4\pi\tau_{Q}q^{2}})\cos\psi\right\rbrace ,
\end{align}
where $\psi$ is the total dynamical phase defined Eq. (\ref{equation_psi}). Then by utilizing the following two integrations $(m=1,~2)$,
\begin{align}
	&\int_{0}^{\pi}\frac{dq}{\pi}\cos(qr)	e^{-2m\pi\tau_{Q}q^{2}}=\sqrt{\frac{m}{\pi\hat{\xi}^{2}}}e^{-mr^2/\hat{\xi}^{2}}, \\
	&\int_{0}^{\pi}\frac{dq}{\pi}e^{-2m\pi\tau_{Q}q^{2}}\cos(qr)\cos\psi	=
	\frac{\sqrt{8}e^{- r^2/\left(l^{\alpha}_{m}\right)^{2}}\sin\phi_{m,r}^{\alpha}}{(2m\pi^{2})^{1/4}\sqrt{\hat{\xi}~l^{\alpha}_{m}}},
\end{align}
where the three lengths, $\hat{\xi}$ and $l^{\alpha}_{m}$'s, and the phases, $\phi_{m,r}^{\alpha}$'s, are defined in Eqs. (\ref{xi}), (\ref{l_alpha}), and (\ref{phi_alpha}), we can get the result in Eq. (\ref{alpha-r}).

For the off-diagonal fermionic correlator, $\beta_{r}$, we first adopt the approximation,
\begin{equation}
  e^{-\pi\tau_{Q}q^{2}}\sqrt{1-e^{-2\pi\tau_{Q}q^{2}}}\approx Y q\sqrt{2\pi\tau_{Q}} e^{-y\pi\tau_{Q}q^{2}}, \label{approximate-beta}
\end{equation}
to make the integral analytically tractable. The two variational parameters, $y$ and $Y$, are to be fixed. This can be done numerically \cite{Dziarmaga_2021}. Here we provide an alternative way, which demands the two sides of Eq. (\ref{approximate-beta}) share the same extremum at their peaks. The left side of Eq. (\ref{approximate-beta}) exhibits a peak at position $q^{*}=\sqrt{\frac{\ln2}{2\pi\tau_{Q}}}$ with the extreme value $1/2$. Meanwhile, the right side of Eq. (\ref{approximate-beta})  exhibits a peak at position $\tilde{q}^{*}=\frac{1}{\sqrt{2\pi y\tau_{Q}}}$ with the extreme value $\frac{Y}{\sqrt{ye}}$. So we get two equations,
\begin{align}
	q^{*}=\tilde{q}^{*},~~ \frac{1}{2}=\frac{Y}{\sqrt{ye}}.
\end{align}
The solutions are $y=\frac{1}{\ln 2}$ and $Y=\frac{1}{2}\sqrt{\frac{e}{\ln 2}}$. Thus we can arrive at

\begin{align}
	u_{q}(t_{f})v_{q}^{*}(t_{f})\approx&\sum_{m=1}^{5}(-1)^{m-1}Y_{m}~qe^{-h_{m}\pi\tau_{Q}q^{2}}\sqrt{2\pi\tau_{Q}}\nonumber\\
	&\times e^{i(\lambda'_{m}+q^{2}\tau_{Q}\lambda_{m})}, \label{C6}
\end{align}
\noindent where $h_{m}$, $Y_{m}$, $\lambda'_{m}$, and $\lambda_{m}$ can be found below Eq. (\ref{beta-r}) in the main text. Next, by substituting Eq. (\ref{C6}) into $\beta_{r}$, we can deduce the integrals like
\begin{align}
  &X_{m}\sqrt{2\pi\tau_{Q}}\int_{0}^{\pi}\frac{dq}{\pi}\sin(qr)qe^{-h_{m}\pi\tau_{Q}q^{2}}e^{i(\lambda'_{m}+q^{2}\tau_{Q}\lambda_{m})}\nonumber\\
  =& \frac{\sqrt{2\tau_{Q}}X_{m}r~e^{i \phi_{m}^{\beta}(r)}}{4\left\lbrace (\pi h_{m}\tau_{Q})^{2}+(\tau_{Q}\lambda_{m})^{2} \right\rbrace^{3/4} }e^{-\frac{\pi h_{m}\tau_{Q}r^{2}/4}{ (\pi h_{m}\tau_{Q})^{2}+(\tau_{Q}\lambda_{m})^{2}}}
\nonumber\\
  =&\frac{y_{m}}{\sqrt{\hat{\xi}l_{m}^{\beta}}}\frac{r}{l_{m}^{\beta}} ~e^{-\left(r/l_{m}^{\beta}\right)^{2} }e^{i \phi_{m}^{\beta}(r)},
\end{align}
where the lengths, $l_{m}^{\beta}$'s, and the phases, $\phi_{m}^{\beta}(r)$, are are defined in Eqs. (\ref{l_beta}) and (\ref{phi_beta}).

\bibliographystyle{aapmrev4-2}

%

\end{document}